\documentclass[11pt]{article}
\usepackage{epsfig}
\usepackage{amssymb,amsmath}
\usepackage{epigraph}
\usepackage{xcolor}

\usepackage[lined,boxed,commentsnumbered]{algorithm2e}

\usepackage{graphicx}				% Use pdf, png, jpg, or epsÃÂ§ with pdflatex; use eps in DVI mode
\usepackage{amssymb}
\usepackage{multirow}

\usepackage{siunitx,booktabs}
\usepackage{newtxtext,newtxmath}

\def\arg{\text{arg}}

\def\boldsigma{\text{\boldmath$\sigma$}}
\def\boldepsilon{\text{\boldmath$\varepsilon$}}

\newcommand{\beq}{\begin{equation}}
\newcommand{\eeq}{\end{equation}}
\newcommand{\bea}{\begin{eqnarray}}
\newcommand{\eea}{\end{eqnarray}}

\begin{document}
%\renewcommand{\thefootnote}{\fnsymbol{footnote}}

%\begin{titlepage}

%\begin{flushright}
% cond-mat/yymmnnn
%\end{flushright}
%\vskip3cm

\begin{center}

{\LARGE
 Phases of MANES: Multi-Asset Non-Equilibrium Skew Model 
\vskip0.5cm
of a Strongly Non-Linear Market with Phase Transitions    

}

\vskip1.0cm
{\Large Igor Halperin\footnote{Fidelity Investments. E-mail: igor.halperin@fmr.com. Opinions expressed here are author's own, and do not represent views of his employer. A standard disclaimer applies. E-mail for communications on the paper: ighalp@gmail.com}  
%and Matthew Dixon\footnote{Department of Applied Math, Illinois Institute of Technology. Email: matthew.dixon@iit.edu}
} \\
\vskip0.5cm
%First version: November 8, 2020 \\
%This version: 
\today \\

\vskip1.0cm
{\Large Abstract:\\}
\end{center}
\parbox[t]{\textwidth}{
This paper presents an analytically tractable and practically-oriented model of non-linear dynamics of a multi-asset market in the limit of a large number of assets.
The asset price dynamics are driven by money flows into the market from external investors, and their price impact.
This leads to a model of a market as an ensemble of interacting non-linear oscillators with the Langevin dynamics.
In a homogeneous portfolio approximation, the mean field treatment of the resulting Langevin dynamics produces the McKean-Vlasov equation as a dynamic equation for market returns. Due to the strong non-linearity
 of the McKean-Vlasov equation, the resulting dynamics give rise to ergodicity breaking and first- or second-order phase transitions under variations of model parameters. Using a tractable potential of the Non-Equilibrium Skew (NES) model previously suggested by the author for a single-stock case, the new Multi-Asset NES (MANES) model enables an analytically tractable framework for a multi-asset market.  The equilibrium expected market log-return is obtained as a self-consistent mean field of the McKean-Vlasov equation, and derived in closed form in terms of parameters that are inferred from market prices of S\&P 500 index options.   
 The model is able to accurately fit the market data for either a benign or distressed market environments, while using only a single volatility parameter.
  
 }
 
 \newcounter{helpfootnote}
\setcounter{helpfootnote}{\thefootnote} 
\renewcommand{\thefootnote}{\fnsymbol{footnote}}
\setcounter{footnote}{0}
\footnotetext{
I would like to thank  John Dance, Lisa Huang, Andrey Itkin, Sebastian Jaimungal, Tal Kachman and Yinsen Miao for comments and helpful remarks. 
}     

 \renewcommand{\thefootnote}{\arabic{footnote}}
\setcounter{footnote}{\thehelpfootnote} 

\newpage
 
\section{Introduction}

When financial practitioners and academics talk about the behavior of the market, they normally refer to the price behavior of stock market indexes such 
as the S\&P 500 or the Dow Jones index. For both these market indexes (and other similar ones), returns are defined as weighted averages of returns of their constituent stocks, the differences being in the chosen universe of stocks and the weighting scheme. In particular, the S\&P 500 index weighs all stocks by their total market capitalization, and is therefore heavily influenced by mega-stocks.

%Obviously, differences in weighting schemes only matter because different stocks are different. In an imaginary world where all individual stocks in an investment universe are identical, a uniform weighting scheme would be the only logically reasonable one. Of course, in reality the stock market is highly heterogeneous in its constituent stocks in all possible senses, and therefore a particular weighting scheme might have a heavy influence on the resulting statistical patterns of price fluctuations of the correspondingly defined index. 

Given that a market index such as the S\&P 500 (the SPX index) is a weighted average of individual stock prices, one might be tempted to 
assume that the dynamics of market returns would be similar to dynamics of a single `representative' stock in the market.
% perhaps with parameters different from those that would apply to any individual stock. 
 Such interchangeability of modeling returns for the whole market versus modeling returns for single stocks is commonly assumed by both academics and practitioners alike. In particular, in both the theory and practice of derivatives pricing, the same models such as stochastic volatility models are used for either single stocks or market indexes, albeit with different parameters.

However, the dynamics of the mean return of all stocks (i.e. the market return) can only be simply expressed as the mean of dynamics of individual returns if these dynamics are \emph{linear}.
In a more general case, we may think of a market as an ensemble of individual stocks whose individual dynamics are generally \emph{non-linear} due to market friction effects as discussed in more details below. Furthermore, these non-linear dynamics for individual stocks are not independent of each other (again, specific market mechanisms producing such co-dependencies will be presented below). Therefore, in general the market dynamics should be viewed as statistical mechanics of 
an interacting ensemble of self-interacting nonlinear 'particles' representing individual stocks. Quantities such as market returns would be computed in such a framework 
as ensemble averages. 

For non-linear interacting systems, ensemble averages do \emph{not} in general reduce to some sort of expectations within a single 
particle dynamics. Therefore, a relation between dynamic properties of a markets index such as SPX and properties of a `typical' or `representative' stock is in general unknown. However, in statistical physics there exists an approach that essentially \emph{constructs} such
effective single-particle dynamics starting with an initial multi-particle interacting system. This approach, or rather a family of approaches, are known in physics as 
\emph{mean field approximations}, or MFA for short. Mean field approximations used in statistical physics are known to become exact in the limit when 
the number $ N $ of particles in the systems goes to infinity, $ N \rightarrow \infty $, known as the thermodynamic limit. As the S\&P 500 index is composed of $ N = 500 $ stocks, such value of $ N $ might be already sufficiently large to make the MFA qualitatively or even quantitatively accurate.          

In this paper, I present a simple and tractable model that builds the dynamics of a market index as the dynamics of the mean log-return of all stocks in the market.
In other words, the market index is identified with the mean field in an ensemble of stocks that compose the market index. 
The way the MFA is constructed and used in this work is similar to how it is used in one of its most canonical applications to the classical Ising model, where the mean field is identified with the mean magnetization of Ising spins, and computed as a solution of a certain self-consistency equation (see e.g. \cite{McQuarrie} or \cite{Zinn-Justin-QFT}).

Similarly, in the model presented below the MFA gives rise to the mean field (i.e. the equilibrium expected market log-return) as a solution of a particular MFA self-consistency equation. 
Differently from the Ising model that deals with binary spins without self-interactions, the model developed below deals with continuous real-valued non-linear (i.e. self-interacting) oscillators as building blocks representing individual stocks. Nonetheless, the mean field approximation applied to the model of this paper similarly produces a self-consistency equation whose solution gives the predicted equilibrium market log-return.   
Moreover, many other details of the model developed in this paper will also bring strong analogies with the Ising model, including the phase structure of the model that admits both first- and second-order phase transitions in different regimes of parameters, and the values of critical exponents.

In this work, the market index portfolio is considered as an ensemble of stocks with individual log-returns $ y_i $ ($ i = 1, \ldots, N $).
They can be viewed as `particles' with `positions' $ y_i $.
The key input to modeling ensembles of interacting particles in statistical physics is a potential function
$ U(y_1, \ldots, y_N) $. The choice of the potential defines all further properties of a model, including both static and dynamic properties. In general, a potential 
$ U(y_1, \ldots, y_N) $ can be decomposed into a sum of self-interaction potentials and interaction potentials. For example, if only pairwise interactions are allowed,
the decomposition takes the following form
\beq
\label{ND_potential_0}
U(y_1, \ldots, y_N) = \sum_{i=1}^{N} V (y_{i}) + \frac{g}{2N} \sum_{i,j=1}^{N} V_{int}(y_i - y_j)
\eeq
where $ V (y_{i}) $ is a self-interaction potential for stock $ i $, $ V_{int}(y_i - y_j) $ is a pairwise interaction potential, and $ g $ is a coupling constant that regulates the strength of interactions in the system.  

In this paper, I motivate the choices for both self-interaction and interaction potentials using the analysis of money flows and their price impact in a multi-asset market. This approach follows the previous work by the author \cite{HD_QED,NES}\footnote{See also \cite{Inverted_World} for a non-technical presentation.} where non-linear models of a single stock price dynamics were obtained starting with a similar analysis of money flows and their impact.\footnote{The critical role of 
money flows and their impact on asset returns were highlighted in recent important papers by Gabaix and Koijen \cite{Gabaix_Koijen_2020} and Bouchaud \cite{Bouchaud_2021}.}
As was shown in \cite{HD_QED}, a combined effect of money flows from external investors and their price impact gives rise to 
a non-linear (cubic) drift $ \mu(x) $ in the diffusion law for the stock price. This translates into stochastic Langevin dynamics where diffusion in the price space is described as a Brownian motion of a particle that is additionally subject to an external non-linear potential $ V(x) $. The latter is defined to satisfy the relation 
$ \mu(x) = - \partial V/ \partial x $, therefore a cubic drift $ \mu(x) $ translates into a quartic polynomial as a model of the potential $ V(x) $.

In \cite{NES}, a closely related model called the Non-Equilibrium Skew (NES) model was presented for the log-return space. 
Unlike more traditional approaches in physics where a potential is an input and stationary or non-stationary (transition) probability distributions are outputs, 
the NES model starts with a parameterized model for a stationary distribution, chosen to be a square of a simple two-component Gaussian mixture. 
The potential in the NES model is given by a negative of a logarithm of this Gaussian mixture. This provides a flexible and highly analytically tractable self-interaction potential $ V(y) $ with five parameters, which can produce either single-well or double-well potentials that have, respectively, either one or two local minima. The outputs of the 
model are transition probabilities for a pre-asymptotic regime, before settling to an equilibrium steady state (which is in fact an input to the model as mentioned above) in the long run. As was shown in \cite{NES}, these pre-asymptotic, non-equilibrium corrections to the asymptotic steady-state return distribution impact estimated moments of the return distribution such as variance, skewness and kurtosis - which explains the name of NES model.         
      
This paper presents a multi-asset extension of the NES model, to be referred to as the MANES model.\footnote{
While the NES model developed in \cite{NES} is a single-stock model, its performance was explored in \cite{NES} using options on the S\&P 500 indexes
rather than single stocks. In this paper, the single-stock model 
proposed in \cite{NES} will be properly used as a building block of a multi-asset MANES model.}  In this framework, the NES potential serves as a self-interaction potential $ V(y) $ in Eq.(\ref{ND_potential_0}), while the interaction potential $ V_{int}(y_i - y_j) $ is found to be quadratic $ V_{int}(y) = \frac{1}{2} y^2 $. The MANES model can be viewed as a new statistical mechanics model with a highly tractable potential that can describe different dynamics 
depending on the model parameters. In particular, the self-interaction potential in the MANES model can be of a double-well form, depending on model parameters. In the latter case, the model behavior is similar to the Desai-Zwanzig model of interacting double-well anharmonic oscillators \cite{Desai_Zwanzig_1978}.
         
As will be shown in detail below, the approach developed in this paper offers a number of insights. First, it links the dynamics of  
market returns with the dynamics of an equivalent single stock that arises in the mean field approximation to the multi-particle dynamics of a market made of $ N $ stocks. Second, the MFA approximation applied to a multi-particle interacting system with the potential (\ref{ND_potential_0}) produces a non-linear extension of the classical Fokker-Planck equation called the McKean-Vlasov equation \cite{Dawson_1983}. Because the McKean-Vlasov equation is non-linear, it leads to ergodicity breaking and phase transitions \cite{Dawson_1983}. As the mean field approximation employed in the McKean-Vlasov equation is accurate in the 
thermodynamical limit $ N \rightarrow \infty $, this suggests that the dynamics of the market index can be successfully modeled as the mean field dynamics for a large ensemble of particles/stocks.

This approach is able to produce a large set of dynamics scenarios. In addition to a benign market regime of small fluctuations around some stationary or time-varying deterministic level (trend), the model also admits regimes of large fluctuations 
involving both first- and second-order phase transitions that can be realized in different parameter regimes. 

Furthermore, I show how the model can be used in practice by calibrating it to market quotes on options on market indexes such as S\&P 500 (SPX) options.
Using a homogeneous approximation, the mean field potential arising with the McKean-Vlasov equation describing an ensemble of non-linear oscillators with the NES self-interaction potentials can be represented as an effective \emph{single-stock} NES potential with parameters that are modified (`renormalized') by interactions. This provides the aforementioned missing link between the dynamics of the market index and the dynamics of a `representative' stock that mimics the index.
When parameters of this effective single-stock NES potential are inferred from the market options data, they are used to compute the equilibrium mean field of the MANES model. The latter is interpreted as a model-based, option-implied prediction of the equilibrium log-return of the market, and can be used as 
a signal driving asset allocation decisions. Furthermore, other moments of the index return distribution inferred from market option quotes can also be used as predictive signals that can be used for investment decisions.
    
The paper is organized as follows. Sect.~\ref{sect_factor_dynamics} gives the derivation of non-linear stochastic dynamics of a multi-asset market that is driven by money flows and their impact. These dynamics are then reformulated as multi-particle Langevin and Fokker-Planck dynamics in Sect.~\ref{sect_market_as_stat_system}. Sect.~\ref{sect_MANES_potential} introduces the NES potential as a tractable approximation to a non-linear self-interaction potential
obtained with the approach of Sect.~\ref{sect_factor_dynamics}. Sect.~\ref{sect_McKean_Vlasov} provides a derivation of the McKean-Vlasov equation - a non-linear version of the Fokker-Planck equation for a multi-particle system that arises within a mean field approximation. Sect.~\ref{sect_McKean_Vlasov_bifurcations} derives the self-consistency equations resulting from using the NES potential within the McKean-Vlasov equation, and then
explores the phase structure of the model including both first- and second-order phase transitions in different parameter regimes, and computes critical exponents. 
The next Sect.~\ref{sect_Experiments} derives a closed-form relation for the equilibrium expected log-return of the market in terms of parameters obtained by calibration to index options, and considers examples of calibration to the market data on the SPX options. The final Section~\ref{sect_Summary} concludes.

\section{Nonlinear stochastic dynamics of market returns}
\label{sect_factor_dynamics}
 
 \subsection{Asset price dynamics with money flows and price impact}
 \label{sect_asset_price_dynamics}
 
 % replaced 'derived' with more verbose language.
 % Our QED model can be 'derived' as followed \cite{IHIF}
 
%\blue{Following \cite{IHIF}, we define the QED model}.

%We adopt the notation and assumption of the portfolio model suggested by Boyd {\it et. al.} \cite{Boyd_2017}.

Let $ {\bf x}_t $ with components 
$  x_{it}  $ be a vector of  asset values with $ i = 1, \ldots, N $ in an investment universe of $ N $ assets 
at the beginning of the period $ [t, t+\Delta t] $, where $ t $ is the current time and $ \Delta t $ is a time step size.    
Let us start with a discrete-time asset price dynamics in the presence of outside investors described by the following equations: 
\bea
\label{r_t_one_more}
&& {\bf x}_{t+ \Delta t} = (1 + {\bf r}_{t} \Delta t)  \circ (  {\bf x}_t  +  {\bf a}_t  \circ  {\bf x}_t \Delta t )  \nonumber, \\
&& {\bf r}_{t}  = r + {\bf w} {\bf z}_t  + {\bf f} ({\bf a}_t)  + \frac{1}{ \sqrt{ \Delta t}} \boldepsilon_t, 
\eea
%where $ \circ $ stands  for an element-wise (Hadamard) product.   
where $ \circ $ stands for the element-wise product.
Here the first equation in  (\ref{r_t_one_more}) defines the change of asset values in the time step $ [t, t+\Delta t] $ as a composition of two changes to their time-$t$ values $ {\bf x}_t $. 
First, at the beginning of the interval, investors adjust positions in each stock $ i $ by buying (or selling, depending on the sign of $ a_{it} $) the amount  $ a_{it} x_{it} \Delta t  $ of the stock
 (so that the action vector $  {\bf a}_t $  is defined as an instantaneous rate of change with the dimension of inverse time). Therefore, immediately after that the asset value is deterministically changed to $ {\bf x}_t^{+} := {\bf x}_t +  {\bf a}_t \circ  {\bf x}_t  \Delta t $.
% by choosing control  $ {\bf u}_t = ({\bf x}_{t}^{+} - {\bf x}_t)/ (\Delta t) $.
% which is applied by the portfolio manager at time $ t $. 
After that, the new portfolio $ {\bf x}_t^{+}  = {\bf x}_t  + {\bf a}_t  \circ  {\bf x}_t \Delta t $ grows at rate $ {\bf r}_{t+\Delta t} $. The latter is given by the second of Eqs.(\ref{r_t_one_more})
that defines the vector of equity returns as a combination of  
% where the term $ {\bf f} ({\bf a}_t) $ describes the price impact of the trade $ {\bf a}_t $. 
a risk-free rate $ r $, predictors  $ {\bf z}_t $ with weights $ {\bf w}$, a multivariate noise $  \boldepsilon_t \sim \mathcal{N} (\cdot| {\bf 0}, {\bf \Sigma}) $,
and a vector-valued market impact factor $  {\bf f} ({\bf a}_t) $ with components 
$ f_i(a_{ti}) $. A particular form for function $  {\bf f} ({\bf a}_t) $ will be presented in the next section, while in this section I proceed with a general form of this function.\footnote{The impact function $  {\bf f} ({\bf a}_t) $ may also depend on other state variables, however we will neglect such additional dependencies, see the next section for more details.}

%Note that the latter is a function of the total trade amount per a unit time step.
%As the first asset $ x_{t0} $ is risk-free, we 
%set $ {\bf w}_{0 \cdot} = 0 $, $ f_0(a_0) = 0 $ and $ \boldepsilon_{t0} = 0 $.
%This term is included because our setting assumes   
%a large institutional-size asset portfolio whose trades $ {\bf a}_t  \Delta t $ may be large enough to move the market. It might be convenient to write the return in
%the second of Eqs.(\ref{r_t_one_more}) as a sum of a term that does not depend on $  {\bf a}_t  $ and a $  {\bf a}_t  $-dependent term:
%\beq
%\label{r_decomp}
% {\bf r}_{t+ \Delta t}  =  {\bf r}_{t+ \Delta t}^{(0)} + {\bf f} ({\bf a}_t),  \; \; \;   {\bf r}_{t+ \Delta t}^{(0)} :=   r + {\bf w} {\bf z}_t  +  \frac{1}{ \sqrt{ \Delta t}} \boldepsilon_t,
%\eeq

Equations (\ref{r_t_one_more}) can also be written for increments $ \Delta  {\bf x}_{t} = {\bf x}_{t + \Delta t} - {\bf x}_{t} $:
\beq
\label{x_incr}
\Delta  {\bf x}_{t} =   {\bf x}_{t}  \circ \left( r + {\bf w} {\bf z}_t  + {\bf f} ({\bf a}_t)  +  {\bf a}_t   \right) \Delta t +    {\bf x}_{t}   \circ \sqrt{ \Delta t}\boldepsilon_t 
\eeq
where terms $ \sim \left( \Delta t \right)^2 $ are omitted assuming that the time step $ \Delta t $ is small enough to justify a continuous-time limit 
$ \Delta t \rightarrow 0 $. 
%Using Eq.(\ref{r_decomp}), the increment $ \Delta  {\bf x}_{t} $ can be written as follows:
%\beq
%\label{x_incr_r0}
%\Delta  {\bf x}_{t} =  {\bf x}_{t}  \circ  {\bf r}_{t+ \Delta t}^{(0)} \Delta t  +  {\bf x}_{t}  \circ \left( {\bf a}_t + {\bf f} ({\bf a}_t)  \right) \Delta t 
%\eeq
In the strict limit $ \Delta t = d t \rightarrow 0 $, Eq.(\ref{x_incr}) transforms into the following stochastic differential equation (SDE):
\beq
\label{SDE}
d  {\bf x}_{t} =   {\bf x}_{t}  \circ \left( r + {\bf w} {\bf z}_t  + {\bf f} ({\bf a}_t) +  {\bf a}_t \right)  d t +    {\bf x}_{t}   \circ  {\boldsigma} \, d {\bf W}_t
\eeq 
where $ {\bf W}_t $ is a standard $ N$-dimensional Brownian motion, and $ \boldsigma $ is a $ N \times N $ volatility matrix such that $ \boldsigma \boldsigma^T =  {\bf \Sigma} $. 

Next consider a vector $ {\bf y}_t $ of period-$T$ log-returns $ y_{it} $ with $ i = 1, \ldots, N $, defined as follows 
\beq
\label{y_it}
y_{it} = \log \frac{x_{it}}{x_{i,t-T}}
\eeq 
Using It{\^o}'s lemma, we obtain the SDE for the log-return vector $ {\bf y}_t $:
\beq
\label{SDE_y}
d  {\bf y}_{t} =   \left( r + {\bf w} {\bf z}_t  + {\bf f} ({\bf a}_t) +  {\bf a}_t  -  \frac{\dot{\bf x}_{t-T}}{{\bf x}_{t-T}} 
- \frac{1}{2} \text{Tr} \, \Sigma \right)  d t +    \boldsigma \, d {\bf W}_t
\eeq 
where $ \dot{\bf x}_{t-T} $ stands for the derivative of the time-lagged asset price $ {\bf x}_{t-T} $ with respect to the calendar time $ t $.
Note that without control $ {\bf a}_t $ and the price impact $  {\bf f} ({\bf a}_t) $, Eq.(\ref{SDE}) becomes a standard multivariate lognormal process with the growth rate given by $ r +  {\bf w} {\bf z}_t $, while Eq.(\ref{SDE_y}) becomes a multivariate normal process with a time-dependent drift due to the term 
$  \dot{\bf x}_{t-T}/ {\bf x}_{t-T}  $. 

On the other hand, Eq.(\ref{r_t_one_more}) or its continuous-time limits (\ref{SDE}) or (\ref{SDE_y}) describe a \emph{controlled} system where
investors buy or sell stocks at rates $ a_i $ at each time step. If investors are rational or bounded-rational, their decisions $ a_{it} $ would depend on the previous
performance of assets.  The problem of finding \emph{optimal} control $ {\bf a}_t $ from the viewpoint of market investors could be further formalized by
specifying their utility (reward) functions, and then solving corresponding Bellman or Hamilton-Jacobi-Bellman (HJB) equations. Instead of following this route, in this paper I employ a more phenomenological way, and choose a simple functional form
of the optimal investment rate $ {\bf a}_t $ as a function of state variables based on general arguments. 
%with non-vanishing controls $ {\bf a}_t \neq 0 $ and price impact $  {\bf f} ({\bf a}_t)  \neq 0$, Eq.(\ref{SDE}) 
%may produce very different dynamics of the state vector $ {\bf x}_t $. 
%Note that for a fixed value of  $ {\bf a}_t $ that does not depend on $ {\bf x}_t $,
%the resulting dynamics could be obtained from the uncontrolled dynamics with  $ {\bf a}_t = 0 $ by a constant shift of the drift and a shift by a deterministic process. 
%However, in reality the decision variables $ {\bf a}_t $ provide a feedback control, and therefore they in general depend on the current state $ {\bf x}_t $ or on its change $ \Delta {\bf x}_{t-1} =  {\bf x}_{t} - {\bf x}_{t-1} $, plus possibly other factors (e.g. alpha signals).

To this end, our specification of investment rates $ {\bf a}_t $ should encode some simple stylized facts about retail investors.
In particular, they usually buy stocks when prices go up (i.e. log-returns  $ {\bf y}_t $ are positive), and sell when prices go down and log-returns are negative. 
This can be formalized by specifying a parametric function $ {\bf a}_t = {\bf a} ({\bf y}_t) $ where a possible dependence on other state variables can be encoded in parameters of this function.
Assuming 
for simplicity that their 
actions are perfectly asymmetric with respect to such scenarios, it produces asymmetry relations $ {\bf a}( - {\bf y}_t ) = - {\bf a} ( {\bf y}_t) $ that 
should be imposed on 
all admissible functional specifications $ {\bf a} ({\bf y}_t) $.
In other words,  $  {\bf a} ({\bf y}_t) $ should be an odd function. 

Based on this, the following specification of the investment rate $ a_{it} $ for stock $ i $ as a simple function of log-return $ y_{it} $ will be used in this work:
\beq
\label{a_t_cubic}
a_{it} = \phi_i y_{it} + \lambda_i y_{it}^3 + \frac{\kappa_i}{N} \sum_{j=1}^{N} y_{jt} 
\eeq
where $ \phi_i, \lambda_i \geq 0 $ are parameters capturing, respectively, a linear and non-linear dependence of investors' allocation decisions on 
the log-return of asset $ i $, and parameter $  \kappa_i \geq 0 $ determines how it depends on the average performance of other assets (i.e. performance of the market).
As will be shown below, parameters $ \lambda_i $ control non-linear effects, while parameters $ \kappa_i $ control interactions in the system.
Note that the functional specification (\ref{a_t_cubic}) can be viewed as a low-order Taylor expansion of a general function, where a constant term and a coefficient in front of $ y_{it}^2 $ are set to zero in order to produce an odd function. 

The definition of the investor flow rate given by Eq.(\ref{a_t_cubic}) both 
refines the previous similar specifications suggested in \cite{HD_QED} and \cite{NES} for a single-stock (1D) case, and extends the approach 
suggested in this previous work to the current case of a multi-asset market with $ N $ stocks. Furthermore, the model developed below assumes that $ N $ is large, which seems to be a good assumption if for example we consider the case $ N = 500 $, by the number of constituents in the S\&P 500 index.

\subsection{Price impact model with dumb money}
\label{sect:quadratic_impact} 

To complete the model specification, we need to define the model of market impact $ \bf{f} ( {\bf a}_t) $. First, 
note that on the grounds of dimensional analysis, as both $ {\bf a}_t $ and $ \bf{f} $ have dimensions of rates, it means that in addition to the ratios $ {\bf a}_t $, 
the market impact can only depend on some dimensionless variables such as e.g. a ratio of the asset price $ x_{it} $ to  
average trading volume over a lookback time window $ [t-T, t] $. In what follows, the market impact $ {\bf f} ({\bf a}_t) $ will be modeled as a parameterized function 
of $ {\bf a}_t $ with constant parameters, thus neglecting such possible additional dependencies.\footnote{If needed or desired, such dependencies can be re-installed by making parameters of a model for $ {\bf f} ({\bf a}_t) $ dependent on these variables.}
%we assume that it is a function of a dimensionless variable, $ f_i ( a_{it} x_{it}) = f_i (z_i) $ where 
%\beq
%\label{z_t}
%z_{it} = \frac{ a_{it} x_{it}}{ \bar{V}_{it}}
%\eeq
%and $ \bar{V}_{it} $ stands for an average trade volume of stock $ i $ measured at time $ t $ over a sliding period  $ [t-T, t] $.

We need a functional specification of the price impact function $ f_i (a_i) $ that captures the most important effects of impact of market flows on market prices. 
First, we want to ensure consistency with the presence of momentum in stock prices. As typically  
a good recent performance of a stock leads to an increased demand, in a short run this typically further increases returns of this stock.
Therefore, at least until cumulative flows are small enough, one should expect a positive co-dependence between market flows and asset returns.
 
However, such positive co-dependence will only persist until the stock becomes ``saturated" or ``crowded". ``Crowding" in a stock occurs when many market participants simultaneously hold large positions in this stock. During periods of market downturns or high volatility, 
when many market participants simultaneous unwind or reduce their 
positions in a crowded stock, it creates a further downward price pressure on the stock, producing diminishing or even negative returns on positions in the stock.   
 
To capture both effects discussed above, we need to produce scenarios where money flows initially produce a positive impact on asset returns, 
but switch to a negative impact once cumulative inflows exceed some threshold value. 
Such a model would be consistent with the 'dumb money' effect of \cite{Frazzini_2008} that predicts that an initial flow into a stock should increase expected returns, but a continuous buildup of inflows into the stock (a 'crowding') leads to diminishing long-term returns. 

To capture such saturation effects, a proper impact function should depend on previous inflows. 
%Let $ a_t  \Delta t $ be the dollar value of the trade in stock $ i $  in the current month $ t $, and 
Let $ \bar{a}_{it}^{\tau} $ be the cumulative inflow rate in the stock $ i $ over the last $ \tau $ periods 
excluding the current period:
\beq
\label{bar_u_t}
 \bar{a}_{it}^{\tau}  :=  \sum_{t' = 1}^{\tau}  a_{i,t - t'} 
 \eeq 
 Now consider a simple model that produces an increasing impact for small values of $ a_{it} $ 
until the sum of $ a_{it} $ and $  \bar{a}_{it}^{\tau} $ does not exceed some fixed value $ \hat{a}_i $, and a decreasing impact for larger values: 
\beq
\label{f_with_abs}
f_i(a_i) = - \eta_i \left( \left| a_{it}+ \bar{a}_{it}^{\tau}  - \hat{a}_i  \right| - \left| \bar{a}_{it}^{\tau} -  \hat{a}_i  \right| \right)
\eeq
where $ \eta_i > 0 $ is a parameter. This piece-linear function vanishes at $ a_i = 0 $ and $ a_i = 2 \left( \hat{a}_i - \bar{a}_{it}^{\tau} \right) $, and reaches its maximum
equal to $ f_{i \star} =  \eta_i \left( \left|  \hat{a}_i  - \bar{a}_{it}^{\tau}  \right| \right) $  at $ a_i =  \hat{a}_i - \bar{a}_{it}^{\tau} $. For large values of $ | a_i | $, this functions asymptotically behaves as $ f_i(a_i)  \sim - \eta_i | a_i | $.

While the functional form (\ref{f_with_abs}) produces the desired behavior, it amounts to a non-differentiable function. We can consider a soft relaxation of this 
function that produces a differentiable approximation: 
\beq
\label{f_with_g}
f_{\beta} (a_i) = - \eta_i \left( H_{\beta} (a_i - b_i) - H_{\beta} (- b_i) \right), \; \; \; b_i :=  \hat{a}_i - \bar{a}_{it}^{\tau}
\eeq  
where the function $ H_{\beta}(z) $ is defined as follows:
\beq
\label{g_soft_abs}
H_{\beta}(z) = \frac{1}{\beta} \log \left( \frac{e^{\beta z} + e^{ - \beta z}}{2} \right) =  \frac{1}{\beta} \log \cosh \left( \beta z \right) 
\eeq
Note that this function provides a soft differentiable relaxation of function $ f(z) = | z | $, so that $ H_{\infty} (z) = | z | $. Furthermore, for finite values of $ \beta $,
the function $ H_{\beta}(z) $ is convex with a minimum at $ z = 0 $ and $ H_{\beta}(0) = 0 $. For small values of $ z $, the Taylor expansion of $ H_{\beta}(z) $ produces 
\beq
\label{Taylor_g}
H_{\beta}(z) = \frac{\beta}{2} z^2 + \frac{\beta^3}{12} z^4 + O( z^6) 
\eeq
while for large values of $ | z| $, we have $ H_{\beta} (z) \rightarrow | z| $. Therefore, the impact function (\ref{f_with_g}) produces the following behavior:
\beq
\label{f_behavior}
f_{\beta}(0) = 0, \; \; \; a_{\star} := \arg \max_{z} f_{\beta}(z) = b_{i}, \; \; \; f_{\beta}(a_{\star}) =  \eta_{i}  H_{\beta}( b_i) 
\eeq 
For small values of the argument, we obtain
\beq
\label{f_a_small}
f_{\beta}(a_i) = - \eta_i \left( 
- \beta b_i  \left( 1 + \frac{\beta^2 b_i^2}{3} \right) a_{i} +  \frac{\beta}{2} \left(1  + \beta^2 b_i^2 \right) a_i^2 - \frac{\beta^3 b_i}{3} a_i^3 + 
\frac{\beta^3}{12} a_i^4 + \ldots \right)
\eeq 
and for $ | a_i | \gg 1 $, we have $ f_{\beta} (a_i) = - \eta_i | a_i | $.
%The impact function  (\ref{f_with_g}) can be viewed as a relaxation of the following non-differentiable function:
%\beq
%\label{f_infty}
%f_{\infty} (z_i) = - \eta_i \left(  \left| z_i - b_i \right|  - \left| - b_i \right| \right)
%\eeq  
Both (\ref{f_with_abs}) and its relaxed form (\ref{f_with_g}) describe a concave impact function that vanishes at $ a_i = 0 $, reaches its peak at $ a_i = b_i $, and then decreases and eventually becomes negative for larger values $  a_i \geq 2 b_i $.\footnote{A similar profile could be reached with an inverted parabola function such as $ f(z) = a z (b-z) $ with parameters $ a, b > 0 $, however I choose (\ref{f_with_g}) over such specification as I want to have an asymptotically linear, rather that quadratic, behavior of 
$ f(z) $.} 

Note for what follows the asymptotic behavior of the combination $ a_{it} + f_i(a_{it}) $ that enters the SDE (\ref{SDE_y}):
 \beq
\label{a_plus_f}
 a_{i} + f_i(a_{i})) = 
\left\{ \begin{array}{cc} 
    \left(1 + \eta_i \beta b_i  \left( 1 + \frac{\beta^2 b_i^2}{3} \right) \right) a_{i} -   \frac{\beta \eta_i}{2} \left(1  + \beta^2 b_i^2 \right) a_i^2 
    + \frac{\beta^3 b_i \eta_i}{3} a_i^3  
    + \ldots  , \; \; \; & \text{for} \; \left| a_i \right|  \ll  1  
    \\
 a_i  - \eta_i  \left| a_i \right|, \; \; \;      
     & \text{for} \; \left| a_i \right| \gg 1  \\
\end{array} \right.
\eeq
Using the cubic law (\ref{a_t_cubic}) for the flow rate, this can also be written as a function of $ y_i = y_{it} $:
 \beq
\label{a_plus_f_y}
 a_{i} + f_i(a_{i})) = 
\left\{ \begin{array}{cc} 
   \xi_i y_{i} + \rho_i y_{i}^2 
    +\zeta_i y_{i}^3  + g_i \left( \frac{1}{N} \sum_{j=1}^{N} y_{j} - y_{i} \right)
    + \ldots  , \; \; \; & \text{for} \; \left| y_{i} \right|  \ll  1  
    \\
 (1- \eta_i )  \lambda_i y_{i}^3, \; \; \;      
     & \text{for} \;  y_{i} \rightarrow \infty  \\
(1+ \eta_i )  \lambda_i y_{i}^3, \; \; \;      
     & \text{for} \;  y_{i} \rightarrow - \infty    
\end{array} \right.
\eeq
where I retained only a linear dependence on the 'mean field' $  \frac{1}{N} \sum_{j=1}^{N} y_{j} $, and parameters are defined in terms of previously defined model parameters as follows:
\bea
\label{params_a_f}
&& \xi_i :=  \left( \phi_i + \kappa_i \right) \left(1 + \eta_i \beta b_i  \left( 1 + \frac{\beta^2 b_i^2}{3} \right) \right), \; \; \; 
g_i := \kappa_i  \left(1 + \eta_i \beta b_i  \left( 1 + \frac{\beta^2 b_i^2}{3} \right) \right),\nonumber \\
&& \rho_i := -   \frac{\beta \eta_i}{2} \left(1  + \beta^2 b_i^2 \right) \phi_i^2, \; \; \;  \zeta_i :=  \frac{\beta^3 b_i \eta_i \phi_i^3}{3} + \lambda_i  \left(1 + \eta_i \beta b_i  \left( 1 + \frac{\beta^2 b_i^2}{3} \right) \right) 
\eea
The functional form of dependence on log-returns $ y_i $ found in Eq.(\ref{a_plus_f_y}) will be used in the next section to introduce a nonlinear potential that 
will be then used for analysis of the joint dynamics of all stocks in the market. 
%Prior to doing that, we still need to introduce the concept of the potential itself, which we consider next.

\section{Markets as interacting nonlinear oscillators}
\label{sect_market_as_stat_system}

\subsection{Langevin dynamics of interacting nonlinear oscillators}

%Let $ S_t $ be the value of a market index at time $ t $. A period-$T$ log-return  $ y_t $ is defined as follows 
%\beq
%\label{y_t_S_t_0}
%y_t = \log \frac{S_t}{S_{t-T}}
%\eeq 
%In this paper, we focus on modeling dynamic distributions of $ y_t $ rather than distributions of market value (or a stock price) $ S_t $. The reason
%is that unlike price dynamics which are non-stationary due to a market or stock drift, returns dynamics can be either stationary or non-stationary. If we define dynamics in terms of 
%log-returns $ y_t $ instead of prices $ S_t $, we can differentiate between regimes of stationary vs non-stationary returns, which would be impossible if a model is formulated in terms of prices $ S_t $. As an interplay between these regimes is a key ingredient of analysis in this paper, we explore dynamics of log-returns 
%$ y_t $ rather than prices $ S_t $. On the other hand, as $ y_T = \log(S_T/S_0) $, knowing a time-$T$ distribution of $ y_T $ also gives a distribution of $ S_T $.

Focusing on an idealized market portfolio made of $ N $ identical stocks, let us consider the case when all model parameters $ \xi_i, g_i, \rho_i, {\bf w}_{i \cdot}, h_i $ etc. are the same for all stocks, so that we write them as $ \xi, g, \rho $ etc.\footnote{The homogeneity assumption will be lifted below in Sect.~\ref{sect_partition_function}.}  Furthermore, while parameters defined in Eqs.(\ref{params_a_f}) are generally time-dependent, here I neglect such potential time dependence, and treat them as constants.

The stochastic differential equation (\ref{SDE_y}) that describes the continuous-time stochastic dynamics of any individual stock log-return $  y_{i} = y_{it} $ 
in this setting can 
be represented as an (overdamped) Langevin equation \cite{Langevin}. In the Langevin approach, the diffusion drift is obtained as a negative gradient of a potential function $ U({\bf y}_t) $:
 \beq
 \label{Langevin_return_ND}
 d y_{i} = - \frac{ \partial U ({\bf y}_{t})}{\partial y_{i} } dt + h d W_{it}
 \eeq
where $ h $ is a volatility parameter\footnote{I changed here the notation for the volatility parameters from $ \sigma_i $ to $ h $, as parameters $ \sigma_i $ will be utilized differently below.}, $ W_{it} $ is a standard Brownian motion, and $ U({\bf y}_t) $ is a potential  
that is decomposed into a sum of self-interaction and pair interaction terms:
\beq
\label{ND_potential}
U({\bf y}_t) = \sum_{i=1}^{N} V (y_{i}) + \frac{g}{2N} \sum_{i,j=1}^{N} V_{int}(y_i - y_j)
\eeq
Here $ V( y_{i}) $ stands for a single-stock self-interaction potential, $ g $ is an interaction constant, and $ V_{int}(y_i - y_j) $ is an interaction potential.  
The Langevin equation (\ref{Langevin_return_ND}) describes diffusion of $ N $ identical particles placed in a common potential $ V(y) $, and in addition interacting via a pairwise interaction potential $ V_{int} $. Using Eq.(\ref{ND_potential}), we can also write Eq.(\ref{Langevin_return_ND}) as follows:
% The Langevin equation (\ref{Langevin_return_ND}) can also be written in a form that explicitly includes the self-potential $ V(y) $ and the interaction 
% potential:
 \beq
 \label{Langevin_return_ND_v2}
% d y_{i} = - \left( \frac{ \partial V(y_i)}{\partial y_{i} }  +  g \left(\frac{1}{N} \sum_{j=1}^{N} y_j  - y_i \right) dt + h d W_{it}
 d y_{i} = - \left( \frac{ \partial V(y_i)}{\partial y_{i} }  + \frac{g}{N} \sum_{j=1}^{N} \frac{ \partial V_{int}(y_i - y_j)}{\partial y_i}    \right) dt + h d W_{it}
 \eeq
In our setting, the choices for potentials $ V, V_{int} $ are informed by Eqs.(\ref{SDE_y}) and (\ref{a_plus_f_y}) adapted to a homogeneous portfolio setting
(with the coupling constant $ g $ obtained similarly from Eq.(\ref{params_a_f})). 
In particular, interactions are induced by the third term in the control law (\ref{a_t_cubic}). The interaction potential $ V_{int}(y_i - y_j) $ is therefore given by the Curie-Weiss quadratic potential:
\beq
\label{pair_interaction_pot}
 V_{int}(y_i - y_j) = \frac{1}{2}  \left( y_{i} - y_{j} \right)^2
 \eeq
While the interaction potential $ V_{int} $ in Eq.(\ref{pair_interaction_pot}) is quadratic, the self-interaction potential $ V(y) $ as suggested by 
Eq.(\ref{a_plus_f_y}) should have higher-order 
non-linearities:
 \beq
\label{U_self_potential}
 V(y) = 
\left\{ \begin{array}{cc} 
   - \theta y  - \frac{\xi}{2}  y^2 - \frac{\rho}{3} y^3 
    - \frac{\zeta}{4}  y^4  
    + \ldots  , \; \; \; & \text{for} \; \left| y \right|  \ll  1  
    \\
- (1- \eta)  \frac{\lambda}{4} y^4, \; \; \;      
     & \text{for} \;  y \rightarrow \infty  \\
- (1+ \eta)  \frac{\lambda}{4} y^4, \; \; \;      
     & \text{for} \;  y_{i} \rightarrow - \infty    
\end{array} \right.
\eeq
where $ \theta $ is an additional parameter given by the sum of all $ {\bf a}_t $-independent terms in Eq.(\ref{SDE_y}). Therefore, for small values of log-returns,
the self-potential (\ref{U_self_potential}) behaves as a quartic potential. It also scales as  $ y^4 $ asymptotically for $ | y | \rightarrow \infty $, though with different coefficients from those appearing in its small-$y$ expansion given by the first line in Eq.(\ref{U_self_potential}). 

Instead of literally using Eq.(\ref{U_self_potential}) as the specification of a self-interaction potential $ V(y) $ , in this paper I will use a somewhat different 
potential that, while  retaining the important non-linearity of Eq.(\ref{U_self_potential}), also offers better analytical tractability, as well as fixes some potential issues with Eq.(\ref{U_self_potential}). These issues are related to the asymptotic behavior of the potential
 $ V(y) $ at  $ y \rightarrow \pm \infty $, as will be discussed next.

First, note that the expression (\ref{U_self_potential}) implies that in order to have a confining potential that grows as $ y \rightarrow \infty $, 
we need to have parameter $ \eta > 1 $.\footnote{While scenarios with an unbounded potential as $ y \rightarrow \infty $ may describe market bubbles that can only occur for short periods of time, they will not be pursued in this work.} 
On the other hand, even if we take $ \eta > 1 $, the potential (\ref{U_self_potential})  is unbounded from below 
for large \emph{negative} values $ y \rightarrow - \infty $. Such an unbounded behavior could describe a corporate default or bankruptcy event when the stock price drops to zero, which would be equivalent to the strict limit $ y \rightarrow - \infty $. This behavior of the model is obtained as a direct consequence of the specification of the cubic flow rate (\ref{a_t_cubic}) which produces a strong negative selloff rate for large negative values of log-returns $ y $ which, together with the price impact function $ f(a_t) \propto - \eta | a_t | $, can drive the stock price all the way to zero, or equivalently to the infinite negative log-return $ y \rightarrow - \infty $.  
   
%The problem with potential (\ref{U_self_potential}) taken at its face value  is related to its asymptotic behavior at $ y \rightarrow \pm \infty $. 
%First, consider the behavior at the positive infinity $ y \rightarrow  \infty $. Eq.(\ref{U_self_potential}) indicates that parameter $ \eta $ should satisfy the condition $ \eta > 1 $ in order to prevent an eventual escape to $ y \rightarrow \infty $. 

Note however that for all practical purposes, instead of being identified with price drops to the strict zero level, corporate defaults or bankruptcies may be associated with events when the stock price drops to a very small but a non-zero 
value (e.g. a few cents given the previous price of \$10, or \$1 given the previous price of \$50, etc.). In terms of log-returns $ y $, such events would correspond 
to sudden drops to a large negative value. The infinite value of $ y $ obtained when the stock price hits the exact zero is therefore solely due to singularity of the transformation $ y_t = \log \left(x_{t}/x_{t-T} \right) $ at $ x_t = 0 $. If a small but non-vanishing stock price is used as the default boundary, the range of practically admissible values of $ y $ becomes finite. 

The last observation implies that we can replace a potential that may become unbounded as $ y \rightarrow - \infty $  
with a \emph{confining} (i.e. growing at $ y \rightarrow \pm \infty $) potential where corporate defaults would correspond to events of reaching a certain 
arbitrarily negative but finite threshold $ y_{\star} < 0 $. If the probability of reaching this barrier is small enough to ensure consistency with the market data, further details of model behavior for yet smaller values of $ y $, which could otherwise differentiate between a confining and unbounded potential, would be immaterial for all practical purposes. 

\begin{figure}[ht]
\begin{center}
\includegraphics[
width=80mm,
height=55mm]{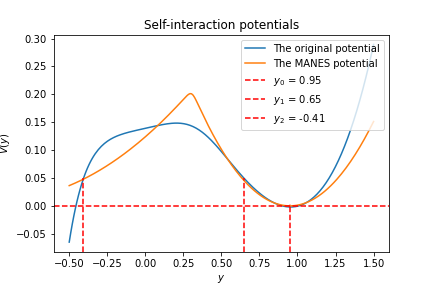}
\caption{Comparison between the  original unbounded potentials $ V(y) $ in Eq.(\ref{U_self_potential}) and the confining MANES potentials (\ref{pot_explicit}). Both the dynamics of local fluctuations around the minimum $ y_0 $ and transition probabilities to move from $ y_1 $ to $ y_2 $ would be similar for the two potentials if potential barriers are sufficiently tall and have similar heights.} 
\label{fig_MANES_vs_original}
\end{center}
\end{figure}

This is illustrated in Fig.~\ref{fig_MANES_vs_original} that shows the original potential $ V(y) $ according to Eq.(\ref{U_self_potential}) alongside a tractable confining potential referred to as the MANES potential, to be presented in details in the next section. Depending on the parameters, the MANES potential can have one or more minima. In Fig.~\ref{fig_MANES_vs_original}, parameters are such that the MANES potential is a double well potential with a second minimum at a negative value of $ y $, which is not shown in Fig.~\ref{fig_MANES_vs_original}.  
The important point to emphasize here is that as long as barriers for both potentials are sufficiently high and have similar heights, both the dynamics of small fluctuations around the potential minimum at $ y_0 $ and probabilities of transitions from $ y_1 $ to $ y_2 $ will be similar in both models. When these probabilities are small, details of behavior in the left tail between the two potentials (an unbounded original potential vs a confining MANES potential) would have a negligible impact on observable consequences of the model. 
 
%While details on this tractable potential  will be given in Sect.~\ref{sect_MANES_potential}.
%we will not need its specific form until then though, as relations presented in Eqs.(\ref{FPE})-(\ref{BBGKY_first}) below and in Sect.~\ref{sect_McKean_Vlasov} are general, and hold for an arbitrary confining self-interaction potential $ V(y) $.   
 
Therefore, instead of literally using Eq.(\ref{U_self_potential}) as a specification of the self-interaction potential $ V(y) $, this work will assume a confining potential
that may be a single-well or a multi-well potential, depending on the model parameters. Confining potentials are easier to work with than non-confining ones, as they give rise to discrete spectra and stationary states in the long run $ t \rightarrow \infty $, which would not exist for non-confining potentials. 
For a multi-well potential with two or more local minima, the model dynamics would be similar to the Desai-Zwanzig model \cite{Desai_Zwanzig_1978} which chooses a symmetric quartic double well potential as a model of $ V(y) $, and its generalized version in 
\cite{Gomes_2019} that considers more complex multi-well potentials. The next section provides details of the MANES potential that will be used later in this paper for the analysis of dynamics and phase structure of the model. 

\subsection{NES and MANES: Multi-Asset dynamics with the Non-Equilibrium Skew potential}
\label{sect_MANES_potential}

%Dynamics implied by the Langevin equation (\ref{Langevin_return_ND_v2}) strongly depend on the topology of the potential $ V(y) $. More specifically, the number of stable states is determined by the number of local minima of the potential. In particular, various non-convex confining potentials of polynomials and piecewise-linear type were considered in \cite{Gomes_2019} to analyze resulting patterns of bifurcations and phase transitions.   

In this paper, the dynamics implied by the Langevin equation (\ref{Langevin_return_ND_v2}) are explored using a simple and highly tractable non-linear self-interaction potential $ V(y) $ given by the logarithm of a two-component Gaussian Mixture (GM):
\beq
 \label{pot_explicit}
  V (y)  = - h^2 \log  \left[ (1-a) \phi(y | \mu_1 T, \sigma_1^2 T) + a  \phi(y | \mu_2 T, \sigma_2^2 T) \right] - h^2 \log C + V_0
  \eeq
 where $ \phi(y | \mu, \sigma^2) $ is a Gaussian density, $ C $ is a constant that will be defined below, and $ V_0 $ is another constant chosen such that the minimum value of $ V(y) $ is zero. This potential was introduced in \cite{NES} within the context of a model for a single stock called the Non-Equilibrium Skew (NES) model. The choice (\ref{pot_explicit}) is motivated by both the ease of the analytical treatment and the flexibility of the parametric family described by Eq.(\ref{pot_explicit}) which, depending on parameters, describes either a single-well anharmonic potential or a double-well potential. As market flows and their price impact produce a nonlinear potential, the parametric family of potentials presented by Eq.(\ref{pot_explicit}) make it possible for  the model to capture these effects. As the present work can be considered a multi-asset (MA) generalization of the NES model, in what follows I will refer to the form in Eq.(\ref{pot_explicit}) as the MANES potential. 
 
\begin{figure}[ht]
\begin{center}
\includegraphics[
width=170mm,
height=110mm]{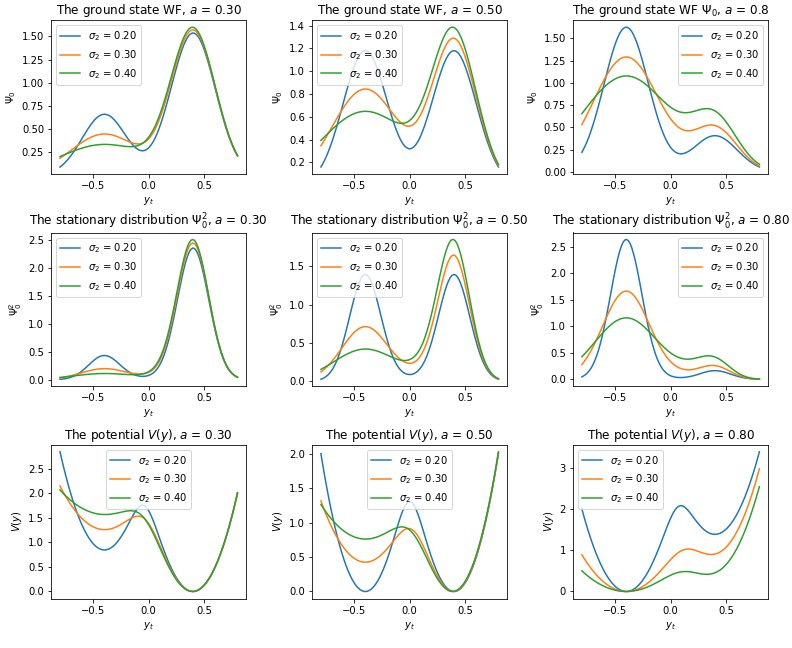}
\caption{The ground state wave function $ \Psi_0(y) $, the stationary distribution $ \Psi_0^2(y) $ and the Langevin potential (\ref{pot_explicit}) as a function of the log-return $ y_t $, for a few values of the asymmetry parameter $ a $, and with different values of parameters $ \sigma_2 $, with fixed values $ \sigma_1 = 0.2,
\mu_1 = 0.4, \mu_2 = - 0.4 $, $ T = 1 $.  Graphs on the left describe a healthy stock where the right minimum is a global minimum. 
Graphs on the right correspond to a severely distressed stock, when the left minimum becomes  a global minimum. Graphs in the middle column describe intermediate scenarios. 
In particular, when $ a = 0.5 $ and $ \sigma_1 = \sigma_2 $, the resulting potential shown in the blue line is symmetric. 
For a multi-asset setting, this potential gives rise to a spontaneous breaking of the $ \mathbb{Z}_2 $ symmetry $ y_t \rightarrow - y_t $.}
\label{fig_IQED_WF_and_potentials}
\end{center}
\end{figure}

 As explained in more detail in \cite{NES}, the MANES potential (\ref{pot_explicit}) can also be represented as 
 \beq
 \label{V_Psi_0}
 V(y) = - h^2 \log \Psi_0(y) + V_0, \; \; \; \Psi_0(y) := C   \left[ (1-a) \phi(y | \mu_1 T, \sigma_1^2 T) + a  \phi(y | \mu_2 T, \sigma_2^2 T) \right] 
 \eeq
  where $ \Psi_0(y) $
 is the ground state wave function (WF) of a quantum mechanical system corresponding to the classical stochastic dynamics with 
 potential $ V(y) $ \cite{NES}.\footnote{A Gaussian mixture can approximate a ground state wave function of a particle placed in either a single well or a double well potential.
 Double well potentials play a special role in statistical physics and quantum mechanics, and are often used to model tunneling phenomena, see e.g. 
 \cite{Landau_QM} or \cite{Zinn-Justin-QFT}. In particular, a symmetric double well is described by a symmetric version of  $ \Psi_0 $ with $ a = 1/2 $ and $ \mu_1 = - \mu_2, \, \sigma_1 = \sigma_2 $.
 For other choices of model parameters, the GM model for function $ \Psi_0 $  can fit a variety of shapes including both a unimodal and bimodal shapes.} 
  In what follows, the WF (\ref{V_Psi_0}) will be occasionally referred to as the MANES WF.
 The stationary state of the original classical stochastic dynamics is then given by its square $ \Psi_0^2(y) $.  The constant $ C $ introduced in (\ref{pot_explicit}) is in fact 
 a  normalization constant that can be obtained from the requirement that the ground state WF $ \Psi_0 $ is squared-normalized, i.e. $ \int dy  \Psi_0^2(y) = 1 $.
 Note that  while  $ \Psi_0(y) $ is proportional to a \emph{two}-component Gaussian mixture, its square is proportional to a \emph{three}-component Gaussian mixture:
 \bea
 \label{three_component_Psi_2}
 \Psi_0^2 (y) &=& \frac{C^2}{2 \sqrt{\pi T}} \left[ \frac{(1-a)^2}{\sigma_1} \phi \left(y | \mu_1 T, \frac{\sigma_1^2}{2} T\right) + 
  \frac{a^2}{\sigma_2} \phi \left(y | \mu_2 T, \frac{\sigma_2^2}{2}  T\right)  \right. \nonumber \\
 & + & \left.
  \frac{ 2 a(1-a)  }{\sqrt{(\sigma_1^2 + \sigma_2^2)/2}} e^{ - \frac{ (\mu_1 - \mu_2)^2 T}{2(\sigma_1^2 + \sigma_2^2)} } 
   \phi \left(y | \mu_3 T, \frac{\sigma_3^2}{2} T \right) \right]
 \eea
  where the additional third Gaussian component has the following mean and variance:
 \beq
 \label{third_component}  
 \mu_3 := \frac{\mu_1 \sigma_2^2 + \mu_2 \sigma_1^2}{\sigma_1^2 + \sigma_2^2}, \; \; \; 
  \frac{\sigma_3^2}{2}  =   \frac{ \sigma_1^2 \sigma_2^2  }{\sigma_1^2 + \sigma_2^2}
 \eeq
 The normalization condition thus fixes the value of the constant $ C $ as follows:
 \beq
 \label{C}
 C^2 =  \frac{2 \sqrt{\pi T}}{ \Omega}, \; \; \; \text{where} \; \; \; \Omega =  \frac{(1-a)^2}{\sigma_1} + 
  \frac{a^2}{\sigma_2}  + 
  \frac{ 2 a(1-a)}{\sqrt{(\sigma_1^2 + \sigma_2^2)/2}} e^{ - \frac{ (\mu_1 - \mu_2)^2 T}{2(\sigma_1^2 + \sigma_2^2)} } 
  \eeq 
% This fixes the value of the constant $ C $ as follows:
% \beq
% \label{C_coeff}
% C^2 =  \frac{2 \sqrt{\pi T}}{ \Omega}, \; \; \; \text{where} \; \; \; \Omega =  \frac{(1-a)^2}{\sigma_1} + 
%  \frac{a^2}{\sigma_2}  + 
%  \frac{ 2 a(1-a)}{\sqrt{(\sigma_1^2 + \sigma_2^2)/2}} e^{ - \frac{ (\mu_1 - \mu_2)^2 T}{2(\sigma_1^2 + \sigma_2^2)} } 
%  \e
The three-component Gaussian mixture density $ \Psi_0^2(y) $ given by Eq.(\ref{three_component_Psi_2}) can be written more compactly using 
Gaussian mixture weights
\beq
\label{GM_weights}
\omega_1 := \frac{(1-a)^2}{\sigma_1 \Omega}, \; \; \; 
\omega_2 := \frac{a^2}{\sigma_2 \Omega}, \; \; \;
 \omega_3 :=   \frac{2 a(1-a)}{ \Omega \sqrt{(\sigma_1^2 + \sigma_2^2)/2}} e^{ - \frac{ (\mu_1 - \mu_2)^2 T}{2(\sigma_1^2 + \sigma_2^2)} }, \; \; \; 
 \sum_{i=1}^{3} \omega_i = 1  
\eeq
This produces the following three-component Gaussian mixture model for the stationary distribution $ p_s(y) = \Psi_0^2(y) $:
 \beq
\label{GM_3_comp}
 p_s(y)  =  \sum_{k=1}^{3} \omega_k \phi \left(y | \mu_k T, \frac{\sigma_k^2}{2} T \right) 
 \eeq
 Examples of trial ground state WFs $ \Psi_0 $ and resulting potentials $ V(y) $ 
 are shown in Fig.~\ref{fig_IQED_WF_and_potentials}.
 
 While this expression produces a non-linear behavior for small positive or negative values of $ y $, its limiting behavior at $ y \rightarrow \pm \infty $ is rather simple and coincides with a harmonic (quadratic) potential:
 \beq
 \left. V(y) \right|_{y \rightarrow - \infty} = h^2 \frac{(y-\mu_2 T)^2}{2 \sigma_2^2 T}, \; \; \; 
 \left. V(y) \right|_{y \rightarrow  \infty} = h^2 \frac{(y-\mu_1 T)^2}{2 \sigma_1^2 T} \; \; \; \; (\mu_2 < \mu_1) 
 \eeq  
The fact that the limiting behavior of the potential coincides with a harmonic potential as $ y \rightarrow \pm \infty $ means that  in this asymptotic regime the model behavior is described by a harmonic oscillator, and thus is fully analytically tractable.

For small values $ y \ll 1 $, the NES potential (\ref{pot_explicit}) can provide a good approximation to a quartic potential that we found for small values $ y \ll 1 $ in the previous analysis, see Eq.(\ref{U_self_potential}). On the other hand, the asymptotic behavior of the NES potential 
(\ref{pot_explicit}) which coincides with a harmonic potential as $ |y| \rightarrow \infty $ is different from the asymptotically quartic potential implied by 
Eq.(\ref{U_self_potential}). However, if the `physics' of the system is determined by a region of small values of $ y $, replacing an asymptotically quartic potential by an asymptotically harmonic potential should produce a negligible impact on observable consequences of the model while significantly simplifying the analysis.  

 As Gaussian mixtures are known to be universal approximations for an arbitrary non-negative functions given enough components, this implies that an \emph{arbitrary} potential that asymptotically coincides with a harmonic oscillator potential can be represented as a negative logarithm of a Gaussian mixture. We can refer to such class of potentials as Log-Gaussian Mixture (LGM) potentials. In this paper, I only consider a two-component LGM potential.\footnote{A requirement of an asymptotic harmonic oscillator behavior could be seen as a potential limitation for the LGM class of potentials, as many interesting potentials
 have a different asymptotic behavior. To this point, we can note that an onset of such a quadratic regime can always be pushed further away by a proper rescaling of the coordinate, while for small or moderate values of a new rescaled argument, the dynamics can still be arbitrarily non-linear, and driven by the number of Gaussian components and their parameters.} 

While with general parameters $ \mu_1 \neq \mu_2, \sigma_1 \neq \sigma_2 $ and $ a \neq 1/2 $ the potential (\ref{V_Psi_0}) is non-symmetric under  
reflections $ y \rightarrow - y $, it becomes symmetric, with $ V(-y) = V(y) $, for the special choice $ \mu_1 = - \mu_2 = \mu $, $ \sigma_1 = \sigma_2 = \sigma $ and $ a = 1/2 $, see Fig.~\ref{fig_IQED_WF_and_potentials}. It is useful for what follows to consider a potential which is only slightly asymmetric. This can be done by considering the following specification of model parameters in (\ref{V_Psi_0}):
\beq
\label{tweaked_params}
a = \frac{1}{2} + \varepsilon_a, \; \; \; \mu_{1,2} = \pm \mu + \varepsilon_{\mu}, \; \; \;  \sigma_{1,2}^2 = \sigma^2 \pm \varepsilon_{\sigma}
\eeq
with small parameters $ \varepsilon_a, \varepsilon_{\mu}, \varepsilon_{\sigma} \ll 1 $. Using these parameters in Eq.(\ref{V_Psi_0}), we obtain, to the linear order in the asymmetry,   
\beq
 \label{V_Psi_0_nonsym}
 V(y) = V^{(s)}(y) - B_0 y
 \eeq
  where $ V^{(s)}(y) = - h^2 \log \Psi_0^{(s)} (y) + V_0 $ is a symmetric potential with $ V^{(s)}(y)  =  V^{(s)}(-y) $, 
  \beq
  \label{Psi_0_sym}
  \Psi_0^{(s)}(y) := \frac{C}{2}   \left[ \phi(y | \mu T, \sigma^2 T) +  \phi(y | - \mu T, \sigma^2 T) \right] 
  \eeq
 is a symmetric ground state wave function, and parameter $ B_0 $ is a linear function of $ \varepsilon_a, \,
 \varepsilon_{\mu}, \, \varepsilon_{\sigma} $:
 \beq
 \label{B_0}
 B_0 =  \frac{h^2}{\sigma^2 T} \left[ \left(1 - \frac{\mu^2}{\sigma^2} \right) \varepsilon_{\mu} 
 + \frac{\mu}{\sigma^2}  \left(1 - \frac{\mu^2}{2 \sigma^2} \right) \varepsilon_{\sigma} \right]  - \frac{2 \mu h^2}{\sigma^2} \varepsilon_a 
 \eeq
 Eq.(\ref{V_Psi_0_nonsym}) shows that a slightly asymmetric potential corresponding to model parameters in Eq.(\ref{tweaked_params}) can be approximated for small values of $ y $ by adding a linear term to the potential $ V^{(s)}(y) $ obtained in the symmetric limit, where the coefficient $ B_0 $ can be directly computed 
 from the original model parameters. As the term $ B_0 y $ in Eq.(\ref{V_Psi_0_nonsym}) can be interpreted as the contribution of an external field $ B_0 $ to the potential energy of the oscillator $ y $, this implies that the dynamics in a slightly asymmetric potentials $ V(y)$ can be approximated by the dynamics in a symmetric potential 
 $ V^{(s)}(y) $ with an additional fictitious external field $ B_0 $.  
 This observation will be used below.

\subsection{The Fokker-Planck equation for MANES}
\label{sect_FP}

Now, after we specified the particular self-interaction MANES potential $ V(y) $ given by Eq.(\ref{pot_explicit}),  we proceed with an equivalent probabilistic 
approach to the dynamics described by the Langevin equation (\ref{Langevin_return_ND_v2}). Such probabilistic method is provided by a corresponding Fokker-Planck equation (FPE). This approach will be presented in the next two sections. Note that in both of them the explicit form of the potential $ V(y) $ is not used, and thus all equations in this section and the next Sect.~\ref{sect_McKean_Vlasov} are general and valid for an arbitrary confining potential $ V(y) $. 

%The stochastic dynamics whose pathwise behavior is encoded in the Langevin equation (\ref{Langevin_return_ND_v2}) can also be described in probabilistic terms using a corresponding Fokker-Planck equation (FPE),  
The FPE corresponding to the Langevin equation (\ref{Langevin_return_ND_v2}) is 
a linear partial differential equation for the joint probability $ P({\bf y},t)  = P( y_{1}, \ldots, y_{N},t) $  
 of a state $ {\bf y} = \left[ y_{1}, \ldots, y_{N} \right] $ describing $ N $ stocks at time $ t $, given an initial position $ {\bf y}_0 $ at time $ t = 0 $: 
 \beq
\label{FPE}
\frac{\partial P({\bf y}, t)}{\partial t} =  \sum_{i=1}^{N} \frac{\partial}{\partial y_i} \left[ \left( \frac{\partial V(y_i)}{\partial y_i} + 
\frac{g}{N} \sum_{j=1}^{N}   \frac{\partial V_{int}(y_i - y_j)}{\partial y_i} 
% g \left(\frac{1}{N} \sum_{j=1}^{N} y_j  - y_i \right) 
 \right)
P({\bf y},t) + \frac{h^2}{2} \frac{\partial}{\partial y_i}  P({\bf y},t ) \right] 
\eeq
For applications, the most interesting probability distributions for a system of identical non-linear oscillators are one-particle density  $ P^{(1)} $ and 
pair-density function $ P^{(2)} $ defined as follows:
\beq
\label{one_particle_dens}
P^{(1)}(y,t) = \int dy_2, \ldots, dy_N P(y, y_2, \ldots, y_N, t), \; \; \; 
P^{(2)}(y,y',t) = \int dy_3, \ldots, dy_N P(y, y', y_3, \ldots, y_N,t)
\eeq
Integrating over $ y_2, \ldots, y_N $ in the FPE (\ref{FPE}), we obtain
\beq
\label{BBGKY_first}
\frac{\partial P^{(1)}( y,t)}{\partial t} =   \frac{\partial}{\partial y} \left[ \frac{\partial V(y)}{\partial y}  P^{(1)}( y,t)  
+ \frac{h^2}{2} \frac{\partial}{\partial y}   P^{(1)}( y,t) +
g \int dy'   \frac{\partial V_{int}(y - y')}{\partial y} 
% g \left(\frac{1}{N} \sum_{j=1}^{N} y_j  - y_i \right) 
P^{(2)}( y,y',t)  \right] 
\eeq
Note that while a single-particle FPE is a partial differential equation (PDE) for a one-particle density, Eq.(\ref{BBGKY_first}) is an integro-differential equation that relates two different densities $ P^{(1)} $ and $ P^{(2)} $. Due to multi-body interactions whose strength is controlled by the coupling constant $ g $, the densities $ P^{(1)} $ and $ P^{(2)} $ are coupled in Eq.(\ref{BBGKY_first}) which in fact represents the first equation in an infinite hierarchy of the BBGKY type, see e.g. \cite{McQuarrie}.  

\subsection{McKean-Vlasov equation for the mean-field dynamics}
\label{sect_McKean_Vlasov}

To proceed, I follow the traditional approach in the physics literature (see e.g. \cite{Martzel_2001}, \cite{Gomes_2019}), and rely on the mean field approximation (MFA) where the probability density factorizes into a product of single-particle densities:
\beq
\label{MFA}
P( y_{1}, \ldots, y_{N},t) = \prod_{i=1}^{N} p(y_i, t) 
\eeq
Therefore, with the MFA, dynamics amounts to a system of $ N $ independent and identical particles, such that the coordinate of any of them 
is given by the average $ \frac{1}{N} \sum_{i=1}^{N} y_{i} $ of all original (interacting) particles in the system. This approximation becomes exact in the thermodynamic limit $ N \rightarrow \infty $.

Note that the mean field $ \frac{1}{N} \sum_{i=1}^{N} y_{i} $ of a homogeneous system of identical non-linear oscillators can be viewed as a reasonable proxy to the market returns, which are usually proxied by returns of the S\&P 500 index. Of course, the real stock market is quite heterogeneous,
and furthermore different firms are weighted in the S\&P 500 index by their total capitalization. Therefore, identifying the mean field $ \frac{1}{N} \sum_{i=1}^{N} y_{i} $ for a homogeneous system with market returns is \emph{not} expected to provide a good approximation for the price dynamics of any individual stock. However, more importantly, the present mean field approach emphasizes the difference between a single-stock dynamics and the group dynamics of a market made of many stocks, while still retaining a link between them and keeping the whole approach practical. 

Plugging the MFA ansatz (\ref{MFA}) into Eq.(\ref{BBGKY_first}), we obtain 
\beq
\label{McKean_Vlasov_0}
\frac{\partial p( y,t)}{\partial t} =   \frac{\partial}{\partial y} \left[ \frac{\partial V(y)}{\partial y}  p( y,t)  
+ \frac{h^2}{2} \frac{\partial}{\partial y}   p( y,t) +
g p(y,t) \frac{\partial }{\partial y}  \int dy'  V_{int}(y - y')  p(y',t)  \right] 
\eeq
This non-linear integro-differential equation holds for an arbitrary interaction potential $ V_{int}(y-y') $. For a particular case of the quadratic Curie-Weiss potential 
(\ref{pair_interaction_pot}), Eq.(\ref{McKean_Vlasov_0}) produces the following equation:
\beq
\label{McKean_Vlasov}
\frac{\partial p( y,t)}{\partial t} =   \frac{\partial}{\partial y} \left[ \frac{\partial V(y)}{\partial y}  p( y,t)  
+ \frac{h^2}{2} \frac{\partial}{\partial y}   p( y,t) +
g  \left(y - \langle y \rangle_{p} \right) p(y,t)  \right], \; \; \; \;  \langle y \rangle_{p} := \int dy y p(y,t)  
\eeq
This equation is non-linear due to the dependence of the coefficient $  \left(y - \langle y \rangle_{p} \right) $ in the last term on the density $ p(y,t) $.
The nonlinear Fokker-Planck equation (\ref{McKean_Vlasov}) is known in the literature as the McKean-Vlasov equation \cite{McKean_1966, Dawson_1983, Frank_2005}. Critically important is the fact the unlike the initial \emph{linear} FPE equation (\ref{FPE}) for a finite $ N $-particle system, the 
the McKean-Vlasov equation that describes the thermodynamic limit $ N \rightarrow \infty $ is \emph{nonlinear}. As a result, it produces far richer dynamics including phase transitions \cite{Dawson_1983}. 
 
Note that if we formally treat $  \langle y \rangle_{p} $ as an independent parameter, the McKean-Vlasov equation (\ref{McKean_Vlasov}) can be
viewed as a linear FPE with the following 'effective' potential:
\beq
\label{pot_eff}
V_{eff}(y) = V(y) + g  \left( \frac{y^2}{2}  - m  y \right), \; \; \; m := \langle y \rangle_p
\eeq
The stationary density for a given value of $ m  = \langle y \rangle_p $ is therefore obtained as a Boltzmann distribution with the `inverse temperature' $ 
\beta = 2/h^2 $:
\beq
\label{Boltzmann}
p(y | m) = \frac{1}{Z(m)} e^{ - \frac{2}{h^2}  \left[ V(y) + g \left(\frac{y^2}{2}  - m  y \right) \right] } , \; \; \; 
Z(m) := \int dy e^{ - \frac{2}{h^2}  \left[ V(y) + g \left(\frac{y^2}{2}  - m  y \right) \right]}
\eeq
This solution should satisfy the self-consistency condition for $ m = \langle y \rangle_p = \int dy y p(y,t) $:
\beq
\label{self_consist}
 m =  \frac{1}{Z(m)} \int dy y e^{ - \frac{2}{h^2}  \left[ V(y) + g \left(\frac{y^2}{2}  - m  y \right) \right]}
\eeq
Once the value of $ m $ is found by solving the self-consistency equation (\ref{self_consist}), it is substituted back to Eq.(\ref{Boltzmann}) to find the stationary density. The number of equilibrium states in the system is therefore given by the number of solutions to Eq.(\ref{self_consist}). 
The solution of this equation for the MANES potential will be presented below in Sect.~\ref{sect_McKean_Vlasov_bifurcations}.
 
\subsection{Renormalization of the single-stock NES potential by interactions}
\label{sect_renormalization}

As was noted above, the McKean-Vlasov equation (\ref{McKean_Vlasov}) can be viewed as a single-particle linear FPE with the effective potential (\ref{pot_eff}) that I repeat here for convenience:
%Note that if we formally treat $  \langle y \rangle_{p} $ as an independent parameter, the McKean-Vlasov equation (\ref{McKean_Vlasov}) can be
%viewed as a linear FPE with the following 'effective' potential:
\beq
\label{pot_eff_2}
V_{eff}(y) = V(y) + g  \left( \frac{y^2}{2}  - m  y \right), \; \; \; m := \langle y \rangle_p
\eeq
provided $ m $ is found from self-consistency equations that would be introduced below. However, prior to this, we can note that 
for the specific choice of the NES potential (\ref{pot_explicit}), the effective potential  $ V_{eff}(y)  $ will have exactly the same functional form as 
the single-stock potential $ V(y) $, albeit with different parameters. Using the physics nomenclature, one can say that parameters of the original single-particle system are \emph{renormalized} by interactions which are represented by the second term in Eq.(\ref{pot_eff_2}). 

To find the new renormalized parameters, we write Eq.(\ref{pot_eff_2}) using Eq.(\ref{pot_explicit}) as follows:
\bea
\label{pot_V_eff_3}
&V_{eff}(y) = - h^2 \log  \left[ (1-a) \phi(y | \mu_1 T, \sigma_1^2 T) + a  \phi(y | \mu_2 T, \sigma_2^2 T) \right]  - h^2 \log e^{ -g \frac{(y-m)^2}{2 h^2}  
+ g \frac{m^2}{2 h^2}} \nonumber \\
& = - h^2 \log \left[ e^{-g \frac{(y-m)^2}{2 h^2}  
+ g \frac{m^2}{2 h^2}} \left(  (1-a) \phi(y | \mu_1 T, \sigma_1^2 T) + a  \phi(y | \mu_2 T, \sigma_2^2 T) \right) \right] \\
& = - h^2 \log  \left[ (1-\bar{a}) \phi(y | \bar{\mu}_1 T, \bar{\sigma}_1^2 T) + \bar{a}  \phi(y | \bar{\mu}_2 T, \bar{\sigma}_2^2 T) \right] - h^2 \hat{V}(m) \nonumber
\eea
where $ \bar{\mu}_1, \bar{\mu}_2, \bar{\sigma}_1, \bar{\sigma}_2, \bar{a} $ are new `renormalized' parameters for a single-particle NES potential, 
and  $ \hat{V}(m) $ stands for terms that depend on $ m $ but not on $ y $. 
They can be easily computed using the well-known relation expressing a product of two Gaussian densities as a rescaled third Gaussian density:
\beq
\label{prod_of_Gaussians}
 \phi(y | \mu_1, \sigma_1^2) \phi(y | \mu_2, \sigma_2^2) = \frac{S_{12}}{\sqrt{2 \pi \sigma_{12}^2 } } \exp\left[ - \frac{(y-\mu_{12})^2}{ 2 \sigma_{12}^2} \right]
 = S_{12} \phi(y | \mu_{12}, \sigma_{12}^2)
 \eeq
 where
 \beq
 \label{prod_GM_params}
 \sigma_{12}^2 = \frac{\sigma_1^2 \sigma_2^2}{\sigma_1^2 + \sigma_2^2}, \; \;  \;
 \mu_{12} = \left( \frac{\mu_1}{\sigma_1^2} + \frac{\mu_2}{\sigma_2^2} \right) \sigma_{12}^2, \; \;  \;
 S_{12} = \frac{1}{ \sqrt{ 2 \pi \left( \sigma_1^2 + \sigma_2^2 \right) }}   \exp\left[ - \frac{(\mu_1 - \mu_{2})^2}{ 2  \left( \sigma_{1}^2 + \sigma_2^2 \right)} \right] 
\eeq
Using these relations, we obtain the renormalized parameters and the function $ \hat{V}(m) $:
\bea
\label{renormalized_params_NES}
& & %\bar{\sigma}_2^2 T = \frac{\sigma_2^2 T}{1 + g \sigma_2^2 T},  \; \;  % \nonumber \\ 
 \bar{\mu}_k  =  \frac{ \mu_k +  \frac{g}{h^2} \sigma_k^2 m }{1 + \frac{g}{h^2}  \sigma_k^2 T}, \; \;  
 \bar{\sigma}_k^2  = \frac{\sigma_k^2 }{1 + \frac{g}{h^2} \sigma_k^2 T}, \; \;  k = 1,2 \nonumber \\ 
%\bar{\mu}_2 T = \left( \frac{\mu_2}{\sigma_2^2} + g m \right) \frac{\sigma_2^2 T}{1 + g \sigma_2^2 T} \nonumber \\
& &\bar{a} = \left[ 1 + \frac{1-a}{a} \sqrt{ \frac{ h^2 + g \sigma_2^2T}{ h^2 + g \sigma_1^2 T}} e^{ \frac{ g (m-\mu_2 T)^2}{2(h^2 + g \sigma_2^2 T)} - 
\frac{g (m-\mu_1 T)^2}{2( h^2 + g \sigma_1^2 T)} } \right]^{-1}  \\ 
& &\hat{V}(m) =  \frac{g m^2}{2 h^2}  + \log \left( 
\frac{ (1-a) h}{\sqrt{h^2+ g \sigma_1^2 T}} e^{ - \frac{ g(m-\mu_1 T)^2}{2(h^2+g \sigma_1^2 T)}  } + 
\frac{ a h }{\sqrt{h^2+ g \sigma_2^2 T}} e^{ - \frac{ g(m-\mu_2 T)^2}{2( h^2+g \sigma_2^2 T)} } 
 \right)  \nonumber   
\eea
These formulae show that additional quadratic and linear terms in Eq.(\ref{pot_eff_2}) can be re-absorbed into rescaled parameters of the Gaussian mixture, where the Gaussian means become linear functions of the mean field $ m $, while the mixing coefficient depends on $ m $ non-linearly. This implies that within the mean field approximation, the  dynamics of market log-returns can be modeled as the dynamics of a fictitious single stock, with initial single-stock model's parameters being 'dressed' by interactions according to Eq.(\ref{renormalized_params_NES}).

\section{Phase transitions of MANES}
\label{sect_McKean_Vlasov_bifurcations}

For a confining potential $ V(y) $, the original finite dimensional Langevin equation (\ref{Langevin_return_ND}) or (\ref{Langevin_return_ND_v2}) produces 
\emph{ergodic and reversible} dynamics under
the Gibbs measure 
 \beq
 \label{Gibbs_measure}
 \mu_N(dy) = \frac{1}{Z_N} e^{- U(y_1, \ldots, y_N)/h^2} dy_1 \cdots dy_N
 \eeq
 where $ Z_N $ is a normalization factor and $ U(y_1, \ldots, y_N) $ is the potential (\ref{ND_potential}), as a consequence of the linearity of the corresponding FPE, and uniqueness of the stationary state of this FPE.
 On the other hand,  the McKean-Vlasov with a confining but non-convex self-interaction potential $ V(y) $ can lead to \emph{violations of ergodicity and phase transitions} \cite{Dawson_1983}.

To explore the phase structure in our setting with the NES potential (\ref{pot_explicit}), first note that using Eq.(\ref{V_Psi_0}), 
the integral entering the self-consistency condition (\ref{self_consist}) can be expressed in terms of an integral that involves $ \Psi_0^2 $:
\beq
\label{self_consist_NES}
 m =  \frac{1}{Z(m)} \int dy y e^{ - \frac{2}{h^2}  \left[ V(y) + g \left(\frac{y^2}{2}  - m  y \right) \right]} = 
 \frac{1}{Z(m)} \int dy y \Psi_0^2 (y)  e^{ - \frac{2g}{h^2} \left(\frac{y^2}{2}  - m  y \right) } 
\eeq
Analysis of the phase structure of the model under variations of parameters is based on Eq.(\ref{self_consist_NES}) where $ m $ is viewed as an order parameter. This is similar to the classical Ising model where the mean magnetization is analogously used as an order parameter, and a second order phase transition occurs at a critical temperature $ T_c $ at which the self-consistency equation $ m = \tanh(m/ T) $ of the Ising model bifurcates and produces a solution $ m \neq 0 $, in addition to the 'trivial' solution $ m = 0 $ which describes a high-temperature regime $ T \rightarrow \infty $.  

A similar pattern of bifurcations and phase transitions can also be found, for certain types of the potential $ V(y) $, for the self-consistency equation 
(\ref{self_consist_NES}) that deals with continuous random variables rather than binary spins of the Ising model. In particular, in the Desai-Zwangiz model,
Eq.(\ref{self_consist_NES}) is solved with the quartic double well potential $ V(y) = - a y^2 + \lambda y^4 $ with $ a, \lambda > 0 $.
However, some general properties of the resulting stochastic system are determined by general properties of the potential, such as the number of local minima, and the asymptotic behavior at $ | y | \rightarrow \infty $, rather than its particular functional form.
In this work I will use the Log-GM NES potential (\ref{V_Psi_0}) which produces a tractable double well potential for certain values of parameters,
however the general analysis below holds for an arbitrary confining potential $ V(y) $.

If the potential $ V(y) $ is symmetric, $ V(-y) = V(y) $, then $ m = 0 $ is always a solution to
the self-consistency equation. This is similar to how the state with zero magnetization is the only stable solution for the Ising model in the high-temperature limit 
(which corresponds to the limit $ h \rightarrow \infty $ in our conventions). However, for lower temperatures, the Ising spins becomes aligned with the mean magnetization $ m = \pm 1 $, producing a bifurcation of the state equation in the parameter space. The bifurcation provides a mean field approximation description of the second order phase transition in the Ising model. Similarly in the current setting, for certain shapes of the potential $ V(y) $ and sufficiently low volatilities $ h $, the self-consistency equation (\ref{self_consist_NES}) can produce multiple solutions below some critical volatility $ h_c $. The analysis of the solution of this equation as a function of the 'temperature' parameter $ h $ should be performed at different values of parameter $ g $ that drives interactions in the system, as well as other parameters of the WF (\ref{V_Psi_0}). The equilibrium second order phase transition describing the order-disorder transitions for the Desai-Zwanzig model was established in \cite{Dawson_1983}. As will be shown next, a similar second-order phase transition can also occur for the MANES model when the potential 
$ V(y) $ is symmetric.

\subsection{Self-consistency equation: second-order phase transition for a symmetric potential}
\label{sect_MANES_self_const}

\begin{figure}[ht]
\begin{center}
\includegraphics[
width=80mm,
height=55mm]{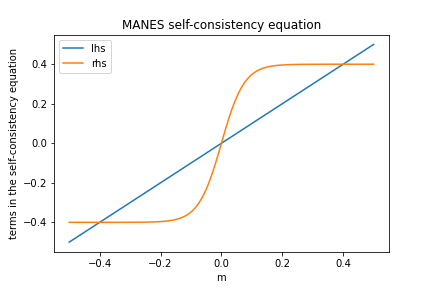}
\caption{The self-consistency equation (\ref{self_consist_NES_3}). The presence of three solutions for the expected log-return $ m $ at $ m = 0 $ and $ m = \pm m_c $ with $ m_c \neq 0 $ is illustrated for the following choice of parameters: $ \mu_1 = - \mu_2 = 0.4, \, \sigma_1 = \sigma_2 = 0.1, \, a = 0.5, \, T = 1.0, \, g = 0.2, \, h = 0.1 $.}
\label{fig_MANES_self_consistency}
\end{center}
\end{figure}

While for many choice of interesting non-convex potentials the analysis requires numerical integration that needs some extra care in the presence of multiple solutions (see \cite{Gomes_2019}), in our case the integral involved in the self-consistency relation (\ref{self_consist_NES}) can be easily computed analytically for
the WF $ \Psi_0 $ defined in Eq.(\ref{V_Psi_0}), with its square given by Eq.(\ref{three_component_Psi_2}). 
%This is done using the well known relation for the product of two Gaussians:
Using Eq.(\ref{prod_of_Gaussians}), we obtain
\beq
\label{self_consist_NES_2}
 m  = \frac{\int dy y \Psi_0^2 (y)    e^{ - \frac{2g}{h^2} \left(\frac{y^2}{2}  - m  y \right) } 
 }{ 
 \int dy   \Psi_0^2 (y)  e^{ - \frac{2g}{h^2} \left(\frac{y^2}{2}  - m  y \right) } 
 }  
 = \frac{h^2}{2g} \frac{\frac{\partial Z(m)}{\partial m}}{Z(m)} = \frac{h^2}{2g} \frac{\partial}{\partial m} \log Z(m)
\eeq
The partition function $ Z(m) $ that enters this expression can be computed using Eqs.(\ref{pot_eff_2}) and (\ref{pot_V_eff_3}):
\beq
\label{Z_m}
Z(m) = \int dy  e^{ - \frac{2g}{h^2}  V_{eff}(y) } = 
e^{ 2 \hat{V}(m) } \left[ \frac{(1-\bar{a})^2}{\bar{\sigma}_1} + \frac{ \bar{a}^2}{\bar{\sigma}_2} + \frac{ 2 \bar{a} ( 1 - \bar{a} )}{ \sqrt{ (\bar{\sigma}_1^2 + 
\bar{\sigma}_2^2)/2}} e^{ - \frac{ (\bar{\mu_1} - \bar{\mu}_2)^2 T}{2  (\bar{\sigma}_1^2 + 
\bar{\sigma}_2^2)}} \right]
\eeq 
where $ \hat{V}(m) $ and parameters $ \bar{a}, \bar{\mu}_1, \bar{\mu}_2 $ etc. are defined in Eq.(\ref{renormalized_params_NES}).
For a symmetric NES potential with $ \mu_1 = - \mu_2 = \mu $, $ \sigma_1 = \sigma_2 = \sigma $ and $ a = 1/2 $, this expression is further simplified:
\beq
\label{Z_m_symm}
Z(m) = \frac{1}{\sigma} e^{2 \hat{V}_s(m) } \left[ 1 - \frac{1 - e^{ - \frac{ h^2 \mu^2 T}{ \sigma^2 (h^2 + g \sigma^2 T)}}}{2 \cosh^2 \left( \frac{g \mu T m}{
h^2 + g \sigma^2 T} \right)} \right]
\eeq
where $ \hat{V}_s(m) $ stands for the symmetric version of $ \hat{V}(m) $ in Eqs.(\ref{renormalized_params_NES}):
\beq
\label{hat_V_s}
\hat{V}_s(m) = \frac{ g m^2}{2 h^2} - \frac{ g (m^2 + \mu^2 T^2)}{2 (h^2 + g \sigma^2 T)} + \log \cosh \frac{g \mu T m}{h^2 + g \sigma^2 T} + \log \frac{h}{
\sqrt{h^2 + g \sigma^2 T}} 
\eeq
The derivative $ \partial \log Z / \partial m $ with the symmetric potential is therefore as follows:
\beq
\label{der_Z_m}
\frac{\partial}{\partial m} \log Z(m) = \frac{ 2 g^2 \sigma^2 T m}{h^2 (h^2 + g \sigma^2 T)} + \frac{ 2 g \mu T}{ h^2 + g \sigma^2 T} 
\frac{ \sinh \frac{ 2 g \mu T m}{h^2 + g \sigma^2 T}}{ \cosh \frac{ 2 g \mu T m}{h^2 + g \sigma^2 T} + e^{ - \frac{ h^2 \mu^2 T}{\sigma^2 ( h^2 + g \sigma^2 T)}}}
\eeq
Substituting this expression into (\ref{self_consist_NES_2}), we obtain the MANES self-consistency equation for a symmetric potential:
\beq
\label{self_consist_NES_3}
m = \mu T \frac{\sinh \left( \frac{2 g \mu T m}{h^2 + g \sigma^2 T} \right) }{
\cosh \left( \frac{2 g \mu T m}{h^2 + g \sigma^2 T} \right) + 
e^{ - \frac{h^2 \mu^2 T}{\sigma^2 (h^2 + g \sigma^2 T )}  } }
\eeq
which looks similar to the self-consistency equation for the Ising model,
%\footnote{especially if the second term in the denominator can be neglected relative to the first term.}
see Fig.~\ref{fig_MANES_self_consistency}.
Note that a symmetric potential assumed in Eq.(\ref{self_consist_NES_3}) may not necessarily match the market data, and actual self-interaction potentials 
$ V(y) $ implied by market prices are typically \emph{not} symmetric (see Sect.~\ref{sect_Experiments}). Nevertheless, analysis of symmetric potentials is of interest because it connects with the theory of second-order phase transitions. As shown below in Sect.~\ref{sect_bifurcations}, for a symmetric potential, the mean field 
vanishes if the volatility $ h $ exceeds a certain critical value $ h_c $, and becomes non-zero, $ m = \pm m_0 $ for some value $ m_0 $, for yet 
lower values $ h < h_c $, with a continuous change from $ m = 0 $ for $ h > h_c $ to non-vanishing values of $ m $ for $ h < h_c $. This describes a continuous (second-order) phase transition, similar to the one obtained for the Ising model with the vanishing magnetic field field ($B=0 $). 

On the other hand, a non-symmetric self-interaction potential $ V(y) $ can be approximated by a symmetric potential $ V^{(s)}(y) $ with a fictitious 'magnetic field' 
$ B_0 $, see Eq.(\ref{V_Psi_0_nonsym}). Driven by the analogy with the Ising model, we could expect that this setting would produce scenarios for a first-order phase transition describing a decay of a metastable state, rather than a second-order phase transition. As we will see next, this is indeed the case in the present model.   

%Note that the potential is chosen to be symmetric in (\ref{self_consist_NES_3}) \emph{not} based on empirical grounds, but simply to produce concise formulas and compare with established models such as the Desai-Zwanzig model and the Ising model. In a more general case of a non-symmetric potential, one should use the more general formula  (\ref{self_consist_NES_2}).
% For a certain parameter range, the resulting behavior of the self-consistent mean field solution is expected to be similar to the one shown in 
%Fig.~\ref{fig_MANES_self_consistency}. 

\subsection{Non-symmetric potential: a first-order phase transition}
\label{sect_first_order_phase_transition}

When the potential $ V(y) $ is not symmetric, the self-consistent mean field $ m $ should be computed using the general formula  (\ref{self_consist_NES_2}).
Such analysis can be further simplified using a linear approximation to
an asymmetric potential by adding a fictitious external field $ B_0 $ to a symmetric potential $ V^{(s)}(y) $, see Eqs.(\ref{V_Psi_0_nonsym}) and (\ref{B_0}). This is equivalent to replacing
 $ m \rightarrow m + \frac{1}{g} B_0 $ and $ V(y) \rightarrow V^{(s)}(y) $ in Eq.(\ref{pot_eff_2}) that defines the effective potential $ V_{eff}(y) $. Therefore, the generalization of 
Eq.(\ref{Z_m_symm}) to the case of a slightly asymmetric potential $ V(y) $ can be obtained by the same replacement $ m \rightarrow m + \frac{1}{g} B_0 $:
 \beq
\label{Z_m_nonsymm}
Z(m) = \frac{1}{\sigma} e^{2 \hat{V}^{(s)} \left( m + \frac{1}{g} B_0 \right) } \left[ 1 - \frac{1 - e^{ - \frac{ h^2 \mu^2 T}{ \sigma^2 (h^2 + g \sigma^2 T)}}}{2 \cosh^2 \left( \frac{g \mu T \left( m + \frac{1}{g} B_0 \right)}{
h^2 + g \sigma^2 T} \right)} \right]
\eeq
Using this expression, we can obtain a generalization of Eq.(\ref{self_consist_NES_3}) for the case $ B_0 \neq 0 $:
\beq
\label{self_consist_NES_4}
m =  \frac{\sigma^2 T }{h^2} B_0 + \mu T \frac{\sinh \left( \frac{2 g \mu T}{h^2 + g \sigma^2 T}   \left( m + \frac{1}{g} B_0 \right) \right) }{
\cosh \left( \frac{2 g \mu T}{h^2 + g \sigma^2 T}  \left( m + \frac{1}{g} B_0 \right) \right) + 
e^{ - \frac{h^2 \mu^2 T}{\sigma^2 (h^2 + g \sigma^2 T )}  } }
\eeq
As implied by this equation, when $ B_0 \neq 0 $, the mean field $ m $ does not vanish for any value of $ h $, and therefore there is no second-order phase transition when $ B_0 \neq 0 $. Instead, in this case we obtain a first-order phase transition under variations of the mean field $ B_0 $. When $ B_0 $ is very 
small but still non-vanishing, its sign determines whether the negative or positive solution with $ m = - m_0 $ or $ m = m_0 $ will have the lowest energy, and 
thus will be the true ground state, where $ \pm m_0 $ are two degenerate mean field solutions obtained in the strict limit $ B_0 = 0 $.  
The discrete $ \mathbb{Z}_2 $ symmetry $ y \leftrightarrow - y $ is explicitly broken when $ B_0 \neq 0 $. If an initial state with e.g. $ m = + m_0 $
is observed at time $ t = 0 $ and the 'magnetic field' $ B_0 > 0 $, then it is the left-well state $ m = - m_0 $ that would be the true ground state, while $ m = + m_0 $
will be a metastable state. Vice versa, for $ B_0 < 0 $, the state $ m = m_0 $ would be the true ground state, while $ m = - m_0 $ could be a metastable state 
released at $ t = 0 $.
A decay of such a metastable state is described as a first-order phase transition, which amounts to a sudden discontinuous jump from 
$ m = - m_0 $ to $ m = m_0 $ (or vice versa, depending on the sign of $ B_0 $ and the initial state). An example of an asymmetric potential leading to such a first-order phase transition will be shown in the next section.
The model thus suggests an interplay between the first- and second-order phase transitions in different regimes of parameters $ (h, B_0 ) $ which is similar to the phase transitions pattern in the Ising model.

\subsection{MANES bifurcation diagrams and phase transitions}
\label{sect_bifurcations}

Analysis of the phase structure of the model is performed using the traditional approach similar to the mean field analysis of the Ising model. We solve the self-consistency equation to compute a function $ m = m(h) $, while keeping other model parameters fixed. The same exercise can be repeated for different values of another important model parameter $ g $ that controls interactions between individual particles (assets).

\begin{figure}[ht]
\begin{center}
\includegraphics[
width=110mm,
height=60mm]{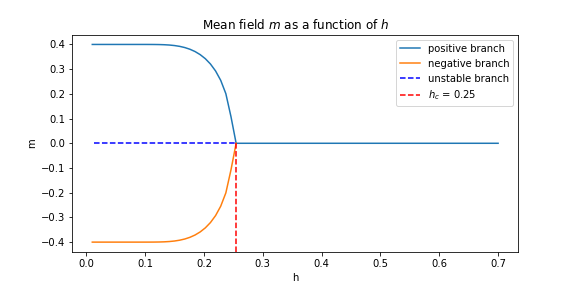}
\caption{Bifurcation diagram of the order parameter $ m $ (the expected market log-return) as a function of the noise level $ h $, obtained 
for a symmetric potential with the following parameters: $ \mu_1 = - \mu_2 = 0.4, \, \sigma_1 = \sigma_2 = 0.1, \,  a = 0.5, \, T = 1.0, \, g = 0.2 $.
The bifurcation occurs at the critical value $ h_c = 0.25 $.
}
\label{fig_MANES_m_vs_h}
\end{center}
\end{figure} 
The phase diagram obtained with this method for a symmetric potential assumed in Eq.(\ref{self_consist_NES_3}) is 
shown in Fig.~\ref{fig_MANES_m_vs_h}. The order parameter $ m $ obtained as a function of a continuously varying volatility parameter $ h $ bifurcates at the critical value $ h_c $. The transition from a zero to a non-zero mean field at the bifurcation point $ h = h_c $ is continuous, as it should be for a second order phase transition.
%and $ h_U $. When $ h_L \leq h \leq h_U $, the order parameter $ m $ 
%does not vanish, while it becomes zero outside of this interval. The model therefore produces two consecutive second order phase transitions occuring, 
%respectively, at two critical value $ h = h_{L} $ and $ h = h_{U} $.

To identify stable versus unstable solutions of the self-consistency equation for either a symmetric or asymmetric potential $ V(y) $ (see, respectively, Eqs.(\ref{self_consist_NES_3}) and (\ref{self_consist_NES_4})), we need to compute the Gibbs free energy $ \mathcal{F} $ which is given by the following expression \cite{Gomes_2019}:
\beq
\label{free_energy}
\mathcal{F} = \frac{h^2}{2} \int p(x) \log p(x) dx + \int V(x) p(x) dx + \frac{g}{2} \int \int V_{int}(x-y) p(x) p(y) dx dy
\eeq
For the stationary density (\ref{Boltzmann}) and the quadratic Curie-Weiss interaction potential, the free energy  
$ \mathcal{F} = \mathcal{F}(m) $ can be computed in a more explicit form:
\beq
\label{free_energy_2}
\mathcal{F}(m) = - \frac{h^2}{2} \log Z(m) + \frac{g}{2} m^2, \; \; \; Z(m) =  
\int dy  \Psi_0^2 (y)  e^{ - \frac{2g}{h^2} \left(\frac{y^2}{2}  - m  y \right) } 
\eeq
The gradient of the free energy with respect to $ m $ is therefore as follows:
\beq
\label{grad_free_energy}
\frac{\partial \mathcal{F}}{\partial m} = - g \frac{ \int dy  \Psi_0^2 (y)  y  e^{ - \frac{2g}{h^2} \left(\frac{y^2}{2}  - m  y \right) }}{
\int dy  \Psi_0^2 (y)  e^{ - \frac{2g}{h^2} \left(\frac{y^2}{2}  - m  y \right) }} + gm 
\eeq
The stationary points of the free energy are obtained by setting this expression to zero, which again produces the self-consistency equation
(\ref{self_consist_NES_2}). 

While the relations (\ref{free_energy}) and (\ref{grad_free_energy}) are general and apply for the generalized Desai-Zwanzig model with an arbitrary self-interaction potential, in this work that uses the log-GM NES potential (\ref{pot_explicit}), the free energy can be computed in closed form 
for either symmetric or asymmetric potentials $ V(y) $ using, respectively Eq.(\ref{Z_m_symm}) or Eq.(\ref{Z_m}). To have even more compact formulae, it is convenient to use the partition function (\ref{Z_m_nonsymm}) corresponding to a weakly asymmetric potential whose asymmetry is controlled by a fictitious external field $ B_0 $. This produces the following expression:
%\bea
%\label{free_energy_non_symmetric_NES}
%\mathcal{F}(m) 
%& = &
%- \frac{g \sigma^2 T}{h^2 + g \sigma^2 T} B_0 m + 
%\frac{ g h^2}{ 2 ( h^2 + g \sigma^2 T) } m^2 
%- h^2 \log \cosh \frac{g \mu T \left( m + \frac{1}{g} B_0 \right)}{ h^2 + g \sigma^2 T} \nonumber \\
%& - & \frac{h^2}{2} \log \left[1 - \frac{1 - b}{2 \cosh^2 \left( \frac{g \mu T \left( m + \frac{1}{g} B_0 \right)}{
%h^2 + g \sigma^2 T} \right)} \right]  + \ldots    \\
%&=& -  \frac{g \sigma^2 T}{h^2 + g \sigma^2 T} B_0 m +  \frac{ g h^2}{ 2 ( h^2 + g \sigma^2 T) } m^2 - 
%\frac{h^2}{2} \log \left[ b + \cosh \left( \frac{2 g \mu T \left( m + \frac{1}{g} B_0 \right) }{ h^2 + g \sigma^2 T} \right) \right] + \ldots   \nonumber 
%\eea
\beq
\label{free_energy_non_symmetric_NES}
\mathcal{F}(m) 
 =  -  \frac{g \sigma^2 T}{h^2 + g \sigma^2 T} B_0 m +  \frac{ g h^2}{ 2 ( h^2 + g \sigma^2 T) } m^2 - 
\frac{h^2}{2} \log \left[ b + \cosh \left( \frac{2 g \mu T \left( m + \frac{1}{g} B_0 \right) }{ h^2 + g \sigma^2 T} \right) \right] + \ldots  
\eeq
where the ellipses stand for omitted constant terms that do not depend on  $ m $, and parameter $ b $ is defined as follows:
\beq
\label{param_b}
b := e^{ - \frac{ h^2 \mu^2 T}{ \sigma^2 (h^2 + g \sigma^2 T)}}
\eeq
Clearly, the free energy $ \mathcal{F}(m) $ is not symmetric, i.e. $ \mathcal{F}(m) \neq \mathcal{F}(-m) $ for $ B_0 \neq 0 $ as $ \mathcal{F}(m) \sim B_0 m + O(m^2) $ for small values of $ m $. More generally, as can be seen from Eq.(\ref{free_energy_2}), the partition function $ Z(m)$ and the free energy $ \mathcal{F}(m)  $ are symmetric as long as $ \Psi_0 (y) $ is symmetric in the $ y$-space, i.e. $ \Psi_0(y) = \Psi_0(-y) $. 
 
\begin{figure}[ht]
\begin{center}
\includegraphics[
width=150mm,
height=65mm]{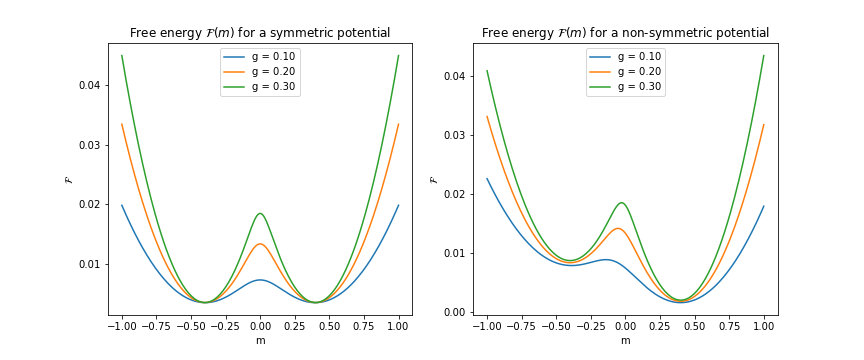}
\caption{Free energy $ \mathcal{F} $ as a function of $ m $. The graph on the left is obtained for a symmetric potential with 
parameters $ \mu_1 = - \mu_2 = 0.4, \, \sigma_1 = \sigma_2 = 0.1, \, a = 0.5, \, T = 1.0$. The graph on the right is obtained for a non-symmetric potential with $  \mu_1 = - \mu_2 = 0.4, \, \sigma_1 = 0.1, \, \sigma_2 = 0.15 $ and $ a = 0.4 $.}
\label{fig_F_vs_m}
\end{center}
\end{figure}

%exact relations for the partition function $ Z(m) $ given by Eq.(\ref{Z_m}) for a general NES potential $ V(y) $, or by Eq.(\ref{Z_m_symm}) for the case of a symmetric potential.
%In particular, for the latter case of a symmetric NES potential we obtain the following expression for the free energy:
%\beq
%\label{free_energy_symmetric_NES}
%\mathcal{F}(m) =  \frac{h^2}{4} \log \left( 1 + \frac{g \sigma^2 T}{h^2} \right) +  \frac{g h^2  \left( m^2 + \mu^2 T^2 \right)}{2 (h^2 + g \sigma^2 T)} - \frac{h^2}{2} \log 
%\left[ \cosh \frac{ 2 g \mu T m}{ h^2 + g \sigma^2 T} + e^{ - \frac{\mu^2 T}{\sigma^2} + \frac{g \mu^2 T^2}{h^2 + g \sigma^2 T}} \right] + \text{const}
%\eeq 
%where a constant stands for for contributions to the logarithm of $ Z $ in Eq.(\ref{Z_m_symm}) that do not depend on 
%$ h$. 
For examples of shapes of the free energy leading to scenarios with phase transitions, see Fig.~\ref{fig_F_vs_m}. Note 
that for the left graph obtained with a symmetric potential with $ \mu_1 = - \mu_2, \, \sigma_1 = \sigma_2, \, a = 0.5 $,
the two minima at $ m = \pm m_c $ with $ m_c \simeq 0.4 $ are degenerate, and the point $ m = 0 $ is a local maximum and is therefore unstable. 
This setting corresponds to the second-order phase transition for $ h < h_c $, and spontaneous breaking of the $ \mathbb{Z}_2 $ symmetry $ m \leftrightarrow - m $ of the free energy $  \mathcal{F}(m)  $ for $ B_0 = 0 $.  
On the other hand, on the right graph, the free energy is non-symmetric for a non-symmetric potential obtained with $ \sigma_1 \neq \sigma_2 $ and $ a \neq 0.5 $,
which can be approximated by having a non-zero field $ B_0 $. For a non-zero field $ B_0 $, the $ \mathbb{Z}_2 $ symmetry $ m \leftrightarrow - m $ of the free energy $  \mathcal{F}(m)  $ is explicitly broken. 
For this case, the potential has a true minimum for a positive value of $ m $ and a local minimum for a negative value of $ m $. If the system is released at time $ t = 0 $ in the state with a negative $ m $, this state is metastable as it has a higher energy than the state with a positive value of $ m $. In this scenario, a transition between the state with $ m < 0 $ and the true ground state with $ m > 0 $ happens very quickly at a random time, and corresponds to a first-order phase transition. 
This behavior is again similar to the Ising model, where adding a non-zero magnetic field breaks the $ \mathbb{Z}_2 $ symmetry of the free energy as a function of magnetization, and produces a first-order phase transition under variations of the temperature.   

\subsection{Critical exponents:  $ \alpha $ and $ \beta $}
\label{sect_critical_exponents}

Critical behavior in the present model is defined in a similar way to the Ising model. For the latter, the critical behavior and bifurcation diagram is usually considered with the temperature being the control parameter and average magnetization being the order parameter, while the external magnetic field is used as an 
additional control parameter. Similarly, in the framework considered here, the volatility parameter $ h $ serves in a similar way to the temperature parameter $ T $ in the Ising model, while the coupling constant $ g $ is used as an additional degree of freedom similar to the magnetic field $ H $ in the Ising model.

To investigate the critical behavior of the model in the vicinity of the phase transition at $ h = h_c $, 
the free energy (\ref{free_energy_non_symmetric_NES}) is expanded into a Taylor series around $ m = 0 $  to the fourth order in $ m $ and second order in $ B_0 $:
\beq
\label{free_energy_small_m}
\mathcal{F}(h^2) =   
- f_1 B_0 m  
+ f_2 m^2  
+ f_4 m^4 + \ldots     
\eeq
where constant terms and higher order terms in $ B_0, m $ are omitted, and parameters $ f_1, f_2, f_4 $ are defined in terms of 
parameter $ b $ introduced in Eq.(\ref{param_b}) and other model parameters as follows:
\bea
\label{params_F_h}
&& f_1  =  \frac{g \sigma^2 T}{ h^2 + g \sigma^2 T}  \left( 1 + \frac{h^2 \mu^2 T^2 (1+b)^{-1} }{h^2 + g \sigma^2 T} \right) \nonumber \\
&& f_2 =  \frac{g h^2}{2 (h^2 + g \sigma^2 T)} \left( 1 - \frac{ 2 g \mu^2 T^2 (1+b)^{-1}}{ h^2 + g \sigma^2 T} \right) \\
&& f_4 = \frac{h^2 (2-b)}{3 (1+b)^2} \left( \frac{g \mu T}{ h^2 + g \sigma^2 T} \right)^4 \nonumber 
\eea
%has the following behavior in different regimes of parameters: 
%\beq
%\label{b_param} 
%b := e^{ - \frac{ h^2 \mu^2 T}{ \sigma^2 (h^2 + g \sigma^2 T)}}
%\simeq
%\left\{ \begin{array}{cc} 
%   e^{ - \frac{h^2 \mu^2}{g \sigma^4}} , \; \; \; & \text{if} \; h^2  \ll  g \sigma^2 T  
%    \\
%  e^{ - \frac{\mu^2 T}{\sigma^2}}, \; \; \;      
%     & \text{if} \;  h^2  \gg  g \sigma^2 T  
%\end{array} \right.
%\eeq
%Note that we obtain $ b < 1 $ for all interesting parameter regimes, therefore for all such regimes the coefficient in front of $ z^4 $ in Eq.(\ref{free_energy_small_m})
%is always positive, ensuring stability of any approximate solution that would be based on the small-$m$ expansion of free energy (\ref{free_energy_small_m}).
The free energy $ \mathcal{F} $ is written in Eq.(\ref{free_energy_small_m}) as a function of the noise variance $ h^2 $ rather than of the mean field $ m $
because in this section we want to explore its dependence on $ h^2 $.

Note that Eq.(\ref{param_b}) implies that $ 0 \leq b \leq 1 $, therefore the coefficient in from of $ m^4 $ is always positive, ensuring stability of 
any approximate solution that would be based on the small-$m$ expansion (\ref{free_energy_small_m}). 
On the other hand, one can see that the coefficient $ f_2 $ can change the sign depending on the value of $ h^2 $.
A bifurcation point $ h^2 = h_c^2 $ corresponds to the value of $ h $ at which the coefficient in 
front of $ m^2 $ in Eq.(\ref{free_energy_small_m}) vanishes, and then becomes negative for yet smaller values $ h^2 < h_c^2 $. This produces the following relation for the critical volatility parameter $ h_c $:
\beq
\label{h_c}
h_c = \sqrt{ 2 g \mu^2 T^2 \left( 1 + b \right)^{-1} - g \sigma^2 T}
\eeq
To produce a real-valued parameter $ h_c $, the expression under the square root should be positive, producing a constraint on parameter combinations that may lead to bifurcation scenarios:
\beq
\label{bifurcation_constraint}
\frac{2 \mu^2 T}{\sigma^2} \geq 1 + b 
\eeq 
The coefficient $ f_2 $ in the expansion (\ref{free_energy_small_m}) can therefore be written in a more suggestive form:
\beq
\label{f_2_coeff_new}
f_2 = \frac{g h^2}{2 (h^2 + g \sigma^2 T)^2} \left( h^2 - h_c^2 \right)
\eeq 
In the vicinity of the bifurcation point $ h = h_c $, the solution for $ m $ in the limit $ B_0 \rightarrow 0 $ can be well approximated by a solution to Eq.(\ref{free_energy_small_m}), which reads
\beq
\label{solution_in_m2_terms}
m^2 = - \frac{f_2}{2 f_4} = \frac{ 3 g (h^2 + g \sigma^2 T)^2}{4 (g \mu T)^4}  \frac{(1+b)^2}{(2-b)} \left( h_c^2 - h^2 \right), \; \; \;  h \leq h_c 
\eeq
%z^2 = \frac{6  \left((1+b)^{-1} - \frac{h^2 + g \sigma^2 T}{2 g \mu^2 T^2}  \right) }{(2-b)(1+b)^{-2}}, \; \; \; 0 \leq \frac{h_c - h}{h_c} \ll 1
%\eeq
This produces the following expression for the mean field 
 $ m $ in the vicinity of the critical point $ h = h_c $: 
\beq
\label{critical_behav}
m = 
%\pm (1+ b) \frac{h^2 + g \sigma^2 T}{2 g \mu^2 T^2} \sqrt{ \frac{6 h_c}{g (2-b)}} \left( h_c - h \right)^{\beta},
\pm \frac{h^2 + g \sigma^2 T}{ 2 (g \mu T)^2} \sqrt{ \frac{ 3 g (1+b)^2}{2-b} }   \left( h_c^2 - h^2 \right)^{\beta}
 \; \; \beta = \frac{1}{2}, \; \; \; \text{if} \; \;  0 \leq \frac{h_c - h}{h_c} \ll 1  
\eeq
which again looks very similar to the relation $ m \sim (T_c - T)^{1/2} $ with the same critical exponent $ \beta = 1/2 $ arising for the Ising model.\footnote{Here $ T $ stands for the temperature, not the time interval as in the previous formulas.}. As in the Ising model, the order parameter $ m $ is continuous across the critical 
point $ h^2 = h_c^2 $, indicating that we deal here with a second-order (continuous) phase transition.

Substituting Eq.(\ref{solution_in_m2_terms}) back into  Eq.(\ref{free_energy_small_m}) gives an approximate expression for the free energy for values of $ h $ that are slightly below the critical value $ h_c $:
\beq
\label{free_energy_small_m_opt}
\mathcal{F}(h^2) =  
% \frac{h^2}{4} \log \left( 1 + \frac{g \sigma^2 T}{h^2} \right) - \frac{g^2 \sigma^2 \mu^2 T^3}{2 (h^2 + g \sigma^2 T)} 
%- \frac{3 h^2}{4}  \frac{ \left( (1+b)^{-1} - \frac{h^2 + g \sigma^2 T}{2 g \mu^2 T^2}  \right)^2}{ (2-b)(1+b)^{-2}}, 
- \frac{f_2^2}{4 f_4} = - \frac{3 h^2}{ 16 g^2 \mu^4 T^4} \left( h_c^2 - h^2 \right)^2 
\; \; \; 0 \leq \frac{h_c - h}{h_c} \ll 1  
\eeq
%The expression (\ref{solution_in_z_terms}) for the mean field $ m $ can be further simplified near the bifurcation point $ h_c $. To this end, we neglect variation of the denominator at approximate it by its value at $ h = h_c $ (i.e. equivalently neglect the $h$-dependence of parameter $ b $ defined in Eq.(\ref{b_param}), and replace it with $ b = b(h_c) $). In addition, we also expand the numerator around its value at $ h_c $. Because the numerator vanishes for $ h = h_c $ due to 
% the definition (\ref{h_c}), the expansion of the numerator starts with a linear term in $ (h - h_c) $. 
%
%We can also simplify the numerator in the last term in Eq.(\ref{free_energy_small_m_opt}) in the same way by expanding it in a first-order Taylor expansion around $ h = h_c $:
%\beq
%\label{free_energy_small_m_opt_2}
%\mathcal{F}(h) =   \frac{h^2}{4} \log \left( 1 + \frac{g \sigma^2 T}{h^2} \right) - \frac{g^2 \sigma^2 \mu^2 T^3}{2 (h^2 + g \sigma^2 T)} 
%- \frac{3 h^2}{4}  \frac{  h_c^2}{ g^2 \mu^4 T^4} \left( h_c - h \right)^2, 
%\; \; \; 0 \leq \frac{h_c - h}{h_c} \ll 1  
%\eeq
While this expression approximates the free energy $ \mathcal{F}(h) $ for $ B_0 = 0 $ for values of $ h $ that are approaching $ h_c $ from below, for values of 
$ h $ that are above $ h_c $, the value of $ \mathcal{F}(h) $ in Eq.(\ref{free_energy_small_m}) will be zero.
%Therefore, we we   
%Using the similarity of our setting with control parameters $ (h, g) $ with the setting of the Ising model with control parameters $ (T,H) $ where $ H $ is an external magnetic field, we can define further critical exponents by analogy with the Ising model. 
%In particular, we 
Further following the analogy with the Ising model, we next define the 'specific heat' $ C_H $ to be proportional to the second derivative of the free energy with respect to noise variance parameter $ h^2 $: 
\beq
\label{spec_heat}
C_H = - h^2 \frac{\partial^2 \mathcal{F} (h^2)}{\partial (h^2)^2} %, \; \; \;  \tau := h^2
\eeq 
Because the expression in Eq.(\ref{free_energy_small_m_opt}) arises only for $ h^2 < h_c^2 $ but vanishes for $ h^2 > h_c^2 $, it is clear that 
the first derivative $ \partial \mathcal{F}(h^2) / \partial h^2 $ is continuous at the critical point $ h^2 = h_c^2 $, but its second derivative
$ \frac{\partial^2 \mathcal{F}(h^2)}{\partial (h^2)^2} $ has a finite jump at $ h^2 = h_c^2 $,  translating into a finite jump of the specific heat at the critical point: 
\beq
\label{jump_CH}
\Delta C_H := - \lim_{\varepsilon \rightarrow 0} \left( \left.  h^2 \frac{\partial^2 \mathcal{F}(h^2)}{\partial (h^2)^2} \right|_{h^2 = h_c^2 - \varepsilon} - 
\left.  h^2 \frac{\partial^2 \mathcal{F}(h^2)}{\partial (h^2)^2} \right|_{h^2 = h_c^2 + \varepsilon} \right) =    \frac{3}{8} \frac{(1+b)^2}{2-b} \frac{h_c^4}{g^2 \mu^4 T^4}
\eeq
This is again the behavior characterizing a second-order (continuous) phase 
transition which is similar to the second-order phase transition of the Ising model. Due to a finite jump, we obtain a vanishing value for the critical exponent $ \alpha $ entering the formula for the specific heat $ C_H \propto | h - h_c|^{\alpha} $, i.e. $ \alpha = 0 $. 
%Thus, the first derivative of the free energy with respect to the noise parameter $ h $ is continues across the critical point $ h = h_c $, but its second derivative experiences a finite jump at this point, again confirming that the model offers a mean field description of a second order phase transition similar to the phase transition in the Ising model. 

\subsection{Partition function and mean field without homogeneity}
\label{sect_partition_function}

To consider other properties of the model such as pairwise correlations, in this section we depart from the approximation of a homogeneous market portfolio that was used above, and consider a heterogeneous market where different stocks may have different parameters $ \mu_i, \sigma_i $.  
The partition function $ Z$ relevant for this setting is given by the following expression:
\beq
\label{Z_partition}
Z \left[ B \right] = \int \prod_{i=1}^{N} d y_i e^{ - \frac{2}{h^2} \left[ \sum_{i=1}^{N}  V_i(y_i) - \sum_{i=1}^{N}  y_i B_i  + \frac{g}{2N} \sum_{i,j=1}^{N} \frac{1}{2} \left( y_i - y_j \right)^2 
 \right] }
\eeq
Here control parameters $ B_i $ are introduced to facilitate calculations of various expectations and correlation functions, and are similar in their meaning to an external magnetic field used in the Ising model and other models of phase transitions for similar purposes.
 
With the Curie-Weiss quadratic interaction potential and MANES self-interaction potential, the partition function can be computed analytically in the limit $ N \rightarrow \infty $. To this end, first note that the interaction term in the potential can be written as follows:
\beq
\label{interaction_term_expand}
e^{ - \frac{2}{h^2}  \frac{g}{2N} \sum_{i \neq j}^{N} \frac{1}{2} \left( y_i - y_j \right)^2  } =  
%e^{  \frac{2}{h^2}  \frac{g}{2N} \sum_{i, j = 1}^{N} y_i  \left( 1 -  \delta_{ij}  \right) y_j   } := 
%e^{  \frac{1}{2} \sum_{i, j = 1}^{N} y_i  J_{ij} y_j    }
e^{ - \frac{2}{h^2}  \left( \frac{N-1}{N} \sum_{i=1}^{N} \frac{1}{2} g y_i^2  - \frac{g}{2 N}  \sum_{i \neq j }^{N} y_i y_j \right)  }
\simeq e^{ - \frac{2}{h^2} \sum_{i=1}^{N} \frac{1}{2} g y_i^2  +  \frac{1}{2} \sum_{i,j=1 }^{N} y_i J_{ij} y_j   }
\eeq
where in the last step I replaced $ (N-1)/N \rightarrow 1 $ assuming that $ N \gg 1 $, and introduced matrix $ {\bf J} $ with matrix coefficients $ J_{ij} = \frac{2 g}{N h^2} \left( 1 - \delta_{ij} \right)  $. This can also be written in the matrix form as follows:
\beq
\label{J_mat}
{\bf J} = \frac{2g}{N h^2} \left( - {\bf I} + {\bf 1} {\bf 1}^{T} \right) 
\eeq   
where $ {\bf I} $ is a unit  $ N \times N $ matrix, and $ {\bf 1} $ is a vector of ones of size $ N $.
%where in the last equation we replaced $ (N-1)/N $ by one, which produces a negligible error in the limit $ N \rightarrow \infty $ assumed here. 
The interaction term can now be represented using integration over auxiliary variables $ \phi_i $ with $ i = 1, \ldots, N $ using
the Hubbard-Stratonovich transformation:
\beq
\label{Hubbard_Stratonovich} 
e^{ \frac{1}{2} \sum_{i,j=1}^{N} y_i  J_{ij} y_j   } = \frac{1}{\sqrt{\text{det} \; J}} \int_{-\infty}^{\infty} \prod_{i=1}^{N} \frac{d \phi_i}{\sqrt{2 \pi}} e^{ - \frac{1}{2} \sum_{i,j = 1}^{N} 
\phi_i  J_{ij} ^{-1}   \phi_j + \sum_{i=1}^{N}  y_i \phi_i }
\eeq
which holds for any real symmetric and invertible matrix $ {\bf J} $, whose inverse matrix has matrix elements $ J_{ij}^{-1} $. 
The inverse of matrix $ {\bf J} $ defined in Eq.(\ref{J_mat}) is computed using the Sherman-Morrison formula:\footnote{The Sherman-Morrison formula  
\[
%\label{Sherman_Morrison}
\left(  {\bf A} + {\bf b} {\bf c}^T  \right)^{-1} = {\bf A}^{-1} - \frac{1}{1 + {\bf c}^{T} {\bf A}^{-1} {\bf b} } {\bf A}^{-1} {\bf b} {\bf c}^{T} {\bf A}^{-1} 
\]
holds for a non-singular matrix $ {\bf A} $ and column vectors $ {\bf b}, \, {\bf c} $ such that the combination $ {\bf A} + {\bf b} {\bf c}^T $ is non-singular.}
\beq
\label{J_inv}
{\bf J}^{-1} =  \frac{N h^2}{2g} \left( - {\bf I} + \frac{{\bf 1} {\bf 1}^{T} }{N-1} \right) 
\eeq
Using Eq.(\ref{Hubbard_Stratonovich}), the partition function (\ref{Z_partition}) can be written as follows:
\beq
\label{Z_second_form}
Z \left[ B \right] =  \frac{1}{\sqrt{\text{det} \; J}} \int_{-\infty}^{\infty} \prod_{i=1}^{N} \frac{d \phi_i}{\sqrt{2 \pi}}  e^{ - \frac{1}{2} \sum_{i,j = 1}^{N} 
\phi_i  J_{ij}^{-1}  \phi_j }
\prod_{i=1}^{N} \int d y_i e^{ - \frac{2}{h^2} \left[ V_i(y_i)  +  \frac{1}{2} g y_i^2  -  \left(\frac{h^2}{2} \phi_i + B_i  \right)  y_i   %  
 \right] }
\eeq
The last expression contains a product of one-dimensional integrals, which can be evaluated in close form by noting that terms proportional to $ y_i $ and $ y_i^2 $ in the exponential can be combined with the potential $ V_i(y_i) $ into a new effective potential similarly to Eqs.(\ref{pot_eff_2}), (\ref{pot_V_eff_3}):
\beq
\label{eff_pot_for_yi}
V_i^{eff}(y_i) := V_i(y_i)   +  g \left( \frac{1}{2} y_i^2  - \psi_i y_i \right), \; \; \; \psi_i :=  \frac{1}{g} \left( \frac{h^2}{2} \phi_i + B_i  \right)     
\eeq
Using Eq.(\ref{pot_V_eff_3}) with $ m $ replaced by $ \psi_i $, we obtain
\beq
\label{inner_product}
\prod_{i=1}^{N} \int d y_i e^{ - \frac{2}{h^2} \left[ V_i(y_i) +  g \left( \frac{1}{2}  y_i^2  - \psi_i  y_i \right)  \right] } 
= \prod_{i=1}^{N} \int d y_i e^{  - \frac{2}{h^2} V_i^{eff}(y_i) } 
 = \prod_{i=1}^{N} e^{ 2 \hat{V}_i(\psi_i) + \log \Omega_i(\psi_i) }
 \eeq
 where function $ \hat{V}_i(\psi_i) $ is defined as in Eq.(\ref{renormalized_params_NES}), and 
%$ \Omega_i (\psi_i)  = \int dy_i \bar{\Psi}_0^2 ( y_i | \psi_i) $ is the normalization factor whose value can be read off Eq.(\ref{C}):
 \beq
 \Omega_i (\psi_i)  = \frac{(1-\bar{a}_i(\psi_i))^2}{\bar{\sigma}_{i1}} +  \frac{ \bar{a}_i^2(\psi_i)}{ \bar{\sigma}_{i2}} + 
 \frac{2 \bar{a}_i (\psi_i) (1 - \bar{a}_i (\psi_i)) }{\sqrt{( \bar{\sigma}_{i1}^2 +  \bar{\sigma}_{i2}^2)/2}} e^{ - \frac{ ( \bar{\mu}_{i1}(\psi_i) - \bar{\mu}_{i2} (\psi_i) )^2 T}{2 ( \bar{\sigma}_{i1}^2 + \bar{\sigma}_{i2}^2 )}}  
 \eeq
 and parameters $ \bar{\mu}_{i1}, \bar{\mu}_{i2}, \bar{\sigma}_{i1}, \bar{\sigma}_{i2}, \bar{a} $ are defined as in Eq.(\ref{renormalized_params_NES}) for each stock $ i $ (I write them here as $ \bar{\mu}_{i1}(\psi_i) $ etc. to emphasizes their dependence on parameters $ \psi_i $).
  
 Substituting (\ref{inner_product}) into Eq.(\ref{Z_second_form}) and changing for convenience the integration variables from $ \phi_i $ to $ \psi_i $ according to 
 Eq.(\ref{eff_pot_for_yi}), the latter can be written as follows:
 \beq
\label{Z_third_form}
Z \left[ B \right] =  \frac{1}{\sqrt{ ( 2 \pi)^N  \text{det} \; J}} \left(\frac{2 g}{h^2} \right)^N 
\int_{-\infty}^{\infty} \prod_{i=1}^{N} d \psi_i  
%e^{ - \frac{1}{2} \sum_{i,j = 1}^{N} 
%\phi_i  J_{ij}^{-1}  \phi_j  + \sum_{i=1}^{N} \left( 2 \log  \hat{V}_i(\psi_i) + \log \Omega_i(\psi_i)  \right) }
e^{ - \frac{2}{h^2} \mathcal{H}({\bf \psi}, B )  }
\eeq
where 
\beq
\label{Lagrangian_for_Z}
\mathcal{H}({\bf \psi}, B ) = \frac{1}{h^2} \sum_{i,j = 1}^{N}   
\left( g \psi_i - B_i \right)  J_{ij}^{-1}  \left(g \psi_j - B_j \right)  - h^2 \sum_{i=1}^{N} \left(  \hat{V}_i(\psi_i) + \frac{1}{2} \log \Omega_i(\psi_i)  \right)
\eeq
is the effective Hamiltonian for variables $ \psi_i $.
The multi-dimensional integral with respect to $ \psi_i $ can be well approximated in the large-$N$ limit $ N \rightarrow \infty $ by a saddle point solution, i.e. 
a solution of the variational equation $ \delta \mathcal{H}({\bf \psi}, B ) / \delta \psi_i = 0 $. 
%corresponding to extrema of function $  \mathcal{H}({\bf \psi}, H ) $ with respect to variables $ \psi_i $. 
The saddle point equation is therefore
\beq
\label{saddle_point_equation}
\frac{2g}{h^2}  \sum_{j=1}^{N} J_{ij}^{-1}   \left( g \psi_j - B_j \right) -  h^2  \left( \frac{ \partial  \hat{V}(\psi_i)}{\partial \psi_i}  +  \frac{1}{2} \frac{ \partial \log \Omega(\psi_i)}{\partial \psi_i}  \right) = 0
\eeq
The solution $ \psi = \bar{\psi} $ of this equation hence satisfies the following equation
\beq
\label{saddle_point_equation_2}
  \bar{\psi}_i  =   \frac{B_i}{g} +  \frac{h^4}{2g^2}  \sum_{j=1}^{N} J_{ij} \left[ \frac{ \partial  \hat{V}(\bar{\psi}_j )}{\partial \bar{\psi}_j}  +  \frac{1}{2} \frac{  \partial \log \Omega(\bar{\psi}_j)}{\partial \bar{\psi}_j} \right]
\eeq
The partition function (\ref{Z_third_form}) with the saddle point approximation then reads
\beq
\label{Z_MF}
Z \left[ B \right] =  \frac{1}{\sqrt{ ( 2 \pi)^N  \text{det} \; J}} \left(\frac{2}{h^2} \right)^N 
e^{ - \frac{2}{h^2} \mathcal{L}( \bar{\psi}, B )  } = e^{ - \frac{2}{h^2} F(B)}
\eeq
where $ \bar{\psi} $ is a solution to Eq.(\ref{saddle_point_equation_2}), and $ F(B) =  \mathcal{H}( \bar{\psi}, B ) $ is the free energy. The local mean field $ m_i  = 
\langle y_i \rangle $ 
is defined as a partial derivative of $ F $:
\beq
\label{m_from_F}
m_i =   \frac{h^2}{2} \frac{ \partial \log Z}{\partial B_i} = - \frac{\partial F}{ \partial B_i} =   \frac{2}{h^2} \sum_{j=1}^{N} J_{ij}^{-1} \left( g \bar{\psi}_j - B_j \right)  
\eeq 
This equation can be inverted to express variables $ \bar{\psi}_i $ in terms of local mean fields $ m_i $:
\beq
\label{psi_from_m}
\bar{\psi}_i =  \frac{1}{g} \left[ \frac{h^2}{2} \sum_{j=1}^{N} J_{ij} m_j + B_i  \right] = \frac{1}{N} \sum_{j \neq i}^{N} m_j + \frac{1}{g} B_i
\eeq
Using Eqs.(\ref{m_from_F}) and (\ref{psi_from_m}), we can now write the saddle point equation (\ref{saddle_point_equation}) in terms of   local mean fields $ m_i $:
\beq
\label{saddle_point_eq_for_m}
m_i =  % h^2  \sum_{j=1}^{N} J_{ij}  \left. 
\left. \frac{h^2}{g} \left[ \frac{ \partial \hat{V}(\bar{\psi}_i )}{\partial \bar{\psi}_i}  +  \frac{1}{2} \frac{ \partial \log \Omega(\bar{\psi}_i)}{\partial \bar{\psi}_i} \right] \right|_{ \psi_i =   \frac{1}{N}  \sum_{j \neq i } m_j  + \frac{1}{g} B_i }
\eeq
The last equation is a general mean field self-consistency equation for the MANES model that defines the local mean field $ m_i $ (i.e. the expectation of the log-return $ y_i $ for the $i$-th stock) in terms of the expected log-returns for other stocks. Other versions of the self-consistency equation can be obtained from 
Eq.(\ref{saddle_point_eq_for_m}) if we make further assumptions. For example, if we use it  with a symmetric single-stock NES potential $ V(y) $,
we obtain  
\beq
\label{saddle_point_eq_for_m_rhs}
  \frac{ \partial \hat{V}(\bar{\psi_i} )}{\partial \bar{\psi_i}}  +  \frac{1}{2} \frac{ \partial \log \Omega(\bar{\psi}_i)}{\partial \bar{\psi}_i}  
% \right|_{ \psi =  \frac{1}{N}  \sum_{j \neq i } m_j  - \frac{1}{g} H_i  } 
  =  \frac{ g^2 \sigma_i^2 T \psi_i }{h^2 (h^2 + g \sigma_i^2 T)} + \frac{ g \mu_i T}{ h^2 + g \sigma_i^2 T} 
\frac{ \sinh \frac{ 2 g \mu_i T \psi_i}{h^2 + g \sigma_i^2 T}}{ \cosh \frac{ 2 g \mu_i T \psi_i}{h^2 + g \sigma_i^2 T} 
+ e^{ - \frac{ h^2 \mu_i^2 T}{\sigma_i^2 ( h^2 + g \sigma_i^2 T)}}} %\right|_{\psi =  \frac{1}{N}  \sum_{j \neq i } m_j  - \frac{1}{g} H_i  } 
\eeq 
For a homogeneous version of this self-consistency equation, we set $ m_i \rightarrow m $, and also remove indices from model parameters 
$ \mu_i \rightarrow \mu, \, \sigma_i \rightarrow \sigma $. Furthermore, we have in this case $ \psi_i \rightarrow \psi = \frac{N-1}{N} m + \frac{1}{g} B $ which is well approximated by $ \psi = m + \frac{1}{g} B $ in the limit of large $ N $.  This produces the self-consistency equation for the homogeneous portfolio:
\beq
\label{saddle_point_eq_for_m_homogeneous}
m =  \frac{\sigma^2 T}{h^2} B + \mu T \frac{ \sinh \left[  \frac{2 g \mu T}{h^2 + g \sigma^2 T}  \left( m + \frac{1}{g} B \right) \right] 
}{
\cosh \left( \frac{2 g \mu T}{h^2 + g \sigma^2 T} \left( m + \frac{1}{g} B \right) \right) + 
e^{ - \frac{h^2 \mu^2 T}{\sigma^2 (h^2 + g \sigma^2 T )}  } }
\eeq
This coincides with Eq.(\ref{self_consist_NES_4}) provided we identify the external control field $ B $ introduced in Eq.(\ref{Z_partition}) with the fictitious field  
$B_0 $ introduced in Eq.(\ref{B_0}) to mimic slightly asymmetric potentials.
 
\subsection{Susceptibilities}
\label{susceptibility}

Formulas for the local mean field approximation developed above enable computing susceptibilities for both the homogeneous and heterogeneous market portfolio settings. Starting with a homogeneous setting, the susceptibility $ \chi $ defined in a similar way to the Ising model:
\beq
\label{suscept_def}
\chi = \left. \frac{ \partial m}{\partial B} \right|_{B \rightarrow 0} 
\eeq
While in the Ising model such expression computes the sensitivity of the mean magnetization to changes of an external magnetic field, in the present setting, $ \chi $ is the sensitivity of the expected market log-return to the amount of asymmetry in the the single-stock self-interaction potential.
 
To compute  the susceptibility $ \chi $,  assume a small but non-vanishing value of $ B $, and expand the right hand side of 
Eq.(\ref{saddle_point_eq_for_m_homogeneous}) to the first order in $ B $. After re-grouping terms, this gives
\beq
\label{m_for_small_B}
m =  \frac{1}{g h^2} \frac{ h^2 h_c^2 + 2 g h^2 \sigma^2 T + g^2 \sigma^4 T^2}{h^2 - h_c^2} B + O \left( B^2 \right)
\eeq
where the critical volatility parameter $ h_c $ is defined in Eq.(\ref{h_c}). This produces the following result for the susceptibility $ \chi $:
\beq
\label{susceptib}
\chi =  \frac{1}{g h^2} \frac{ h^2 h_c^2 + 2 g h^2 \sigma^2 T + g^2 \sigma^4 T^2}{h^2 - h_c^2}
\eeq
Therefore, at the critical volatility value $ h^2 = h_c^2 $, $ \xi $ diverges as $ (h_c^2 - h^2)^{-1} $, similar to the Ising model behavior. 
 
A similar analysis can be performed without using the homogeneous portfolio setting but rather working with Eq.(\ref{saddle_point_eq_for_m}) defined for the local mean fields (i.e. expected values) $ m_i $.  Again, for small external fields $ B_i \rightarrow 0 $, we also expect the local fields $ m_i $ to be small. In this regime, one can retain only the leading linear term in 
$ \psi_i $ in the expansion of the second term in (\ref{saddle_point_eq_for_m_rhs}), and write
\beq
\label{saddle_point_eq_for_m_rhs_linear}
 \left. \frac{h^2}{g} \left[ \frac{ \partial \hat{V}(\bar{\psi_i} )}{\partial \bar{\psi_i}}  +  \frac{1}{2} \frac{ \partial \log \Omega(\bar{\psi}_i)}{\partial \bar{\psi}_i} \right] \right|_{\psi =  \frac{1}{N}  \sum_{j \neq i } m_j  + \frac{1}{g} B_i  } =
A_i \left( \frac{1}{N} \sum_{j \neq i} m_j + \frac{1}{g} B_i \right)  
 \eeq
 where
 \beq
 \label{A_i}
 A_i :=   \frac{ h^2 h_c^2 + 2 g h^2 \sigma_i^2 T + g^2 \sigma_i^4 T^2}{ \left(h^2 + g \sigma_i^2 T \right)^2}
 \eeq
 Plugging Eq.(\ref{saddle_point_eq_for_m_rhs_linear}) back into Eq.(\ref{saddle_point_eq_for_m}), the latter can be written as a linear system of equations
 \beq
 \label{lin_system_for_m}
 \sum_{j} {\bf G}_{ij} m_j =  \frac{1}{g} A_i B_i
 \eeq
 where $ {\bf G} $ is a matrix with matrix elements 
 \beq
 \label{B_ij} 
 G_{ij} = \delta_{ij} \left( 1 + \frac{A_i}{N} \right) - \frac{A_i}{N} 
 \eeq
The solution of (\ref{lin_system_for_m}) is
\beq
\label{lin_system_solution}
{\bf m} =  \frac{1}{g} {\bf G}^{-1} \cdot \left( {\bf A} \circ {\bf B} \right)
\eeq
where $  {\bf A} \circ {\bf B} $ stands for a direct product of vectors $ {\bf A} $ and ${\bf B} $.  This produces the following result for local susceptibilities
\beq
\label{dmdB}
\chi_{ij} := \frac{\partial m_i}{\partial B_j} =  \frac{1}{g} G_{ij}^{-1} A_j
\eeq
The inverse of matrix $ {\bf G} $ can be found using the Sherman-Morrison formula:
\beq
\label{G_inv}
G_{ij}^{-1} = \frac{1}{1 + \frac{A_j}{N}} \delta_{ij} + \frac{1}{1-\frac{1}{N} \sum_{i} A_i } \frac{A_i}{N+A_i} \frac{N}{N + A_j} 
\eeq
In the limit of large $ N $, this can be well approximated by a simpler expression
\beq
\label{G_inv_2}
G_{ij}^{-1} = \delta_{ij} + \frac{1}{1 - \langle A \rangle} \frac{A_i}{N}, \; \; \; \langle A \rangle := \frac{1}{N} \sum_{i=1}^{N} A_i
\eeq

\subsection{Log-return covariances and fluctuation-response relations}
\label{sect_correlations}

The covariance of log-returns for two assets $ i $ and $ j $ is defined as follows:
\beq
\label{log_ret_cov}
C_{ij} := \langle y_i y_j \rangle - \langle y_i \rangle \langle y_j \rangle = \langle \left (y_i - \langle y_i \rangle \right) \left (y_j - \langle y_j \rangle \right) \rangle
\eeq 
This can be computed from the the partition function as follows:
\beq
\label{ret_cov_from_Z}
C_{ij} =  \left. \left(  \frac{h^2}{2} \right)^2 \frac{\partial^2}{\partial B_j \partial B_j } \log Z[B] \right|_{B \rightarrow 0} = 
 \left. \frac{h^2}{2} \frac{ \partial m_i}{\partial B_j} \right|_{B \rightarrow 0}
\eeq
where $ m_i $ is the local mean field defined in Eq.(\ref{m_from_F}). Using Eqs.(\ref{dmdB}) and (\ref{G_inv_2}), we obtain
\beq
\label{C_ij}
C_{ij} = 
\left\{ \begin{array}{cc} 
   \frac{h^2}{2 g N} \frac{A_i A_j}{ 1 - \langle A \rangle } , \; \; \; & \text{if} \; i \neq j  
    \\
  \frac{h^2}{2 g} A_i, \; \; \;      
     & \text{if} \;  i = j  
%- (1+ \eta)  \frac{\lambda}{4} y^4, \; \; \;      
%     & \text{for} \;  y_{i} \rightarrow - \infty    
\end{array} \right.
\eeq
The first relation here shows that in the mean field approach, we obtain $ C_{ij} \propto A_i A_j $ implying that the covariance matrix $ {\bf C} $ has rank one. This is similar to the relation $ C_{ij} \propto \beta_i \beta_j $ that arises in a one-factor model with a common 'market' factor for all stocks, with $ \beta_i $ being the 
regression coefficient of stock $ i $'s return on the market return.  

A link with the calculations performed for the homogeneous setting is provided by considering a special case of the partition function (\ref{Z_partition}) for a homogeneous external field which is the same for all stocks, i.e. $ B_i \rightarrow B $. For this case, the first derivative of $ \log Z $ reads 
\beq
\label{Z_p}
\frac{\partial \log Z}{\partial B} =  \frac{2}{h^2} \frac{1}{Z} \int \prod_{i=1}^{N} d y_i  \sum_{i=1}^{N} y_i 
e^{ - \frac{2}{h^2} \left[ \sum_{i=1}^{N}  V_i(y_i) - \sum_{i=1}^{N}  y_i B  + \frac{g}{2N} \sum_{i,j=1}^{N} \frac{1}{2} \left( y_i - y_j \right)^2 
 \right] }
  =  \frac{2N}{h^2} \langle m \rangle
\eeq
Differentiating once more, one obtains
\beq
\label{Z_pp}
\frac{\partial^2 \log Z}{\partial B^2} =  \left( \frac{2}{h^2} \right)^2 \left[ 
%\left\langle \sum_{i} y_i \sum_{j} y_j \right\rangle - 
\langle \sum_{i} y_i \sum_{j} y_j \rangle -
%\left\langle \sum_{i} y_i \right\rangle \left\langle \sum_{j} y_j \right\rangle 
\langle \sum_{i} y_i \rangle \langle \sum_{j} y_j \rangle 
\right] = 
\left(  \frac{2}{h^2} \right)^2  \sum_{i,j} \text{Cov} \left(y_i, y_j \right) =  \frac{2 N}{h^2} \frac{ \partial \langle m \rangle }{\partial B}
\eeq
where Eq.(\ref{Z_p}) is used at the last step. Re-arranging the last equation here, we obtain the following relation between 
the ``magnetic susceptibility" $ \chi $ and the average 
covariance:
\beq
\label{mean_cov}
\frac{1}{N^2} \sum_{i,j} C_{ij} =  \frac{h^2}{2 N} \frac{ \partial m}{\partial B} =  \frac{h^2}{2 N} \chi
\eeq
where $ \chi $ is computed in Eq.(\ref{susceptib}). This relation shows that the susceptibility is driven by the fluctuations in the system.
Such relations are known in statistical physics as \emph{fluctuation-response} formulae.

For a homogeneous setting with $ A_i = A $, the mean covariance entering Eq.(\ref{mean_cov}) can be computed as follows:
\beq
\label{mean_C_ij}
\bar{C} = \frac{1}{N^2} \sum_{i,j} C_{ij} =  \frac{1}{N^2} \left( \frac{N h^2}{2 g} \frac{A^2}{1-A} + \frac{N h^2}{2 g} A \right) = \frac{h^2}{2gN} \frac{A}{1-A} 
\eeq
Using here Eq.(\ref{A_i}) with $ A_i $ replaced by  $ A$, we finally obtain
\beq
\label{mean_C_2}
\bar{C} =   \frac{1}{2 g N} \frac{ h^2 h_c^2 + 2 g h^2 \sigma^2 T + g^2 \sigma^4 T^2}{ h^2  - h_c^2} =  \frac{h^2}{2 N} \chi
\eeq
where $ \chi $ is the susceptibility for the homogeneous market computed in Eq.(\ref{susceptib}). We have therefore verified that covariances for a heterogeneous market defined in Eq.(\ref{C_ij}) reduce for a homogeneous market portfolio to the average covariance $ \bar{C} $ which is proportional to the susceptibility $ \chi $.

In addition to establishing a correspondence with the homogeneous market setting, the local mean field approach of this section leads to the following observation. While the susceptibility $ \chi $ diverges as $ (h^2 - h_c^2)^{-1} $ as $ h \rightarrow h_c $, this divergence originates in off-diagonal elements $ C_{ij} $ with $ i \neq j $, and arises due to the factor $ 1 - \langle A \rangle $ in the denominator in the first relation in Eq.(\ref{C_ij}). Therefore, while off-diagonal covariances are proportional to $ 1/N $ and are hence parametrically small in the limit of large $ N $, their values are increased as the volatility parameter $ h $ approaches its critical value $ h_c $. This implies that, for certain combinations of parameters that produce scenarios with $ h^2 \simeq h_c^2 $, covariances and/or correlations between different stocks obtained in this modeling framework can be made comparable with the average correlation between stocks in the real market, which is around 0.4 for stocks in the S\&P 500 universe.

%Therefore, to compute covariances $ C_{ij} $, we need to find derivatives $ \partial m_i / \partial H_j $. This can be done using the self-consistency equation (\ref{saddle_point_eq_for_m}). To this end, we first differentiate Eq.(\ref{psi_from_m}) to compute 
%the derivatives  
%\beq
%\label{psi_from_m_der}
%\frac{ \partial \bar{\psi}_i}{\partial H_k} =  \frac{h^2}{2g} \sum_{j=1}^{N} J_{ij} \frac{\partial m_j}{\partial H_k} -   \frac{1}{g} \delta_{ik}
%\eeq

\section{Fitting  model parameters using option data}
\label{sect_Experiments}
 
 As was shown in Sect.~\ref{sect_renormalization}, with the mean field approximation, the dynamics of the market index can be represented as a single-stock dynamics with renormalized parameters $ \bar{a} $ and $  \bar{\mu}_k, \;  \bar{\sigma}_k $ (with $ k = 1,2 $) given by Eq.(\ref{renormalized_params_NES}) which I repeat here for relations involving $  \bar{\mu}_k, \;  \bar{\sigma}_k $:
 \beq
\label{renormalized_params_NES_2} 
 \bar{\mu}_k  =  \frac{ h^2 \mu_k +  g \sigma_k^2 m }{h^2 + g \sigma_k^2 T}, \; \;  
 \bar{\sigma}_k^2  = \frac{ h^2 \sigma_k^2}{h^2 + g \sigma_k^2 T}, \; \;  k = 1,2 
\eeq
Therefore, the resulting effective single-stock model has six parameters $   \bar{\mu}_1,  \bar{\mu}_2,  \bar{\sigma}_1, \bar{\sigma}_2, a , h $. In \cite{NES}, this model specification was further reduced to five parameters by setting $ \bar{\mu}_1 = - \bar{\mu}_2 = \bar{\mu} $. 
%This choice may produce a bias in estimated  expected returns under the real measure, however calibration to option prices involves moments of return distributions under a risk-neutral measure instead of the real measure, and any such bias would \emph{not} be recoverable solely from the option price data that informs the risk-neutral distribution \cite{NES}.
%On the other hand, any such potential bias could be addressed, if needed, with downstream applications of the model. For example, if option-implied signals are used to predict the future return, any systematic bias could be captured by a constant intercept in a linear regression of realized returns on the option signals. 
%This suggests that setting $ \bar{\mu}_1 = - \bar{\mu}_2 = \bar{\mu} $ might be a reasonable choice, especially if the model is calibrated \emph{only} to option prices.
%Yet, to keep the presentation in a more general form, this choice will \emph{not} be assumed below but rather will be kept optional.    
In this paper, such constraint will not be imposed. 

Calibration of the model to market prices of S\&P 500 options (SPX options) or other index options therefore produces six parameters  
$   \bar{\mu}_1,  \bar{\mu}_2,  \bar{\sigma}_1, \bar{\sigma}_2, \bar{a} , h $. 
% possibly reduced to five parameters if we set $ \bar{\mu}_1 = - \bar{\mu}_2 = \bar{\mu} $.
On the other hand, the full set of model parameters in the original multi-asset version of the model involves seven parameters
$ \mu_1, \mu_2, \sigma_1, \sigma_2, a, h, g $, plus one more unknown value of the expected log-return $ m$, which thus effectively serves as the eighth parameter.
Clearly, as eight parameters cannot be uniquely recovered from six parameters $ \bar{\mu}_1,  \bar{\mu}_2,  \bar{\sigma}_1, \bar{\sigma}_2, a , h $ that could be found by calibration to SPX options, we need additional constraints to fix their values. The next few sub-sections develop such constraints on model parameters.

\subsection{Fixing the coupling constant $ g $ and volatilities $ \sigma_k $}
\label{sect_fixing_g}

To estimate the coupling constant $ g $ that controls interactions in the system, one can try to fix it by 
fitting the
single-stock volatility and pair-wise return correlation obtained in our homogeneous portfolio setting to the average stock vol and correlations obtained in the real market. 
This can be readily done using the homogeneous version of Eqs.(\ref{C_ij}). It gives the mean single-stock volatility 
\beq
\label{mean_vol}
\bar{\sigma}_M  =  \frac{h}{ \sqrt{T}} \sqrt{  \frac{A}{2 g} }  =  \frac{h}{ \sqrt{T}}  \sqrt{\frac{ h^2 h_c^2 + 2 g h^2 \sigma^2 T + g^2 \sigma^4 T^2}{ 2 g \left(h^2 + g \sigma^2 T \right)^2}}
\eeq
where Eq.(\ref{A_i}) was used at the last step. Note that the factor $ 1 / \sqrt{T} $ in the right-hand side of this relation arises because $ \bar{\sigma}_M $ is defined in annualized terms.
The mean pairwise correlation is obtained from Eqs.(\ref{C_ij}) as the ratio of the first row to the second row:
\beq
\label{mean_corr}
\bar{\rho}_M =  \frac{1}{N} \frac{A}{1-A}  
\eeq
Note that $ \bar{\rho}_M \sim 1/N $, which implies that correlations die off in the strict thermodynamic limit $ N \rightarrow \infty $. 
The mean correlation $ \rho_M $ can also be estimated differently by computing the variance of the mean field $ m $ in terms of $ \rho_M $ and the mean 
single-stock volatility $ \sigma_M $:
\beq
\label{var_m}
\text{Var}(m) = \left( \frac{1}{N} \bar{\sigma}_M^2 + \bar{\rho}_M \bar{\sigma}_M^2 \right) \Delta t
\eeq 
Neglecting the first term in the right hand side, this relation gives an approximation for $ \rho_M $ in terms of the ratio of annualized variance of the market index' log-return $ \sigma_m^2 $ to  the variance $ \bar{\sigma}_M^2 $ of the representative stock in the portfolio:
\beq
\label{rho_M_est}
\bar{\rho}_M \simeq \frac{ \sigma_m^2}{\bar{\sigma}_M^2} 
\eeq
where $ \sigma_m^2 = \text{Var}(m)/ \Delta t $ is the annualized variance of the market log-return. This produces an estimate $ \bar{\rho}_M \simeq 0.3-0.4 $ which can be used to produce further estimates for parameters in the model. In particular,
inverting Eq.(\ref{mean_corr}) to find $ A $ in terms of $ \bar{\rho}_M $, and then inverting Eq.(\ref{mean_vol}) to compute $ g $, we obtain
\beq
\label{A_and_g}
A = \frac{ N \bar{\rho}_M}{ 1 +  N \bar{\rho}_M}, \; \; \; g 
%= \frac{h^2}{2 \bar{\sigma}_M^2 T}  \frac{ N \bar{\rho}_M}{ 1 +  N \bar{\rho}_M}
= \frac{h^2 A}{2 \bar{\sigma}_M^2 T} 
\eeq 
This suggests that $ A $ should be very close to one, approaching it from below, and respectively $ g \simeq h^2/(2 \bar{\sigma}_M^2 T) $.
Next, we can invert the second relation in Eq.(\ref{renormalized_params_NES_2}) to obtain
\beq
\label{sigma_from_sigma_hat}
\sigma_k^2 T = \frac{ \bar{\sigma}_k^2 T}{ 1 - \frac{g}{h^2} \bar{\sigma}_k^2 T} 
%\simeq \frac{ \bar{\sigma}_k^2 T}{ 1 - \frac{\bar{\sigma}_k^2}{\bar{\sigma}_M^2}}
\eeq
%where at the second step I used the approximation $ g \simeq \frac{h^2}{\bar{\sigma}_M^2 T} $ suggested by Eq.(\ref{A_and_g}). 
This relation shows that $ g $ should be such that $ \frac{g}{h^2} \bar{\sigma}_k^2 T < 1 $ in order to keep the single-stock volatility real-valued and 
finite, $ g $ should satisfy the following constraint:
\beq
\label{g_constraint_0}
g < \frac{h^2}{\bar{\sigma}_k^2 T}
\eeq
Using the second relation in (\ref{A_and_g}), the constraint (\ref{g_constraint_0}) can also be re-stated as  
the following constraint on the volatility $ \sigma_M $ of the `representative' stock:
\beq
\label{constr_sigma_M}
\bar{\sigma}_M^2 \geq \frac{\bar{\sigma}_k^2}{2} A \simeq \frac{\bar{\sigma}_k^2}{2}, \; \; \; k = 1,2
\eeq
where the second approximate form follows as long as $ A \simeq 1 $ as suggested by Eq.(\ref{A_and_g}).
The second observation with Eq.(\ref{sigma_from_sigma_hat}) is that the single-stock variance $ \sigma_k^2 T $ is higher than the variance of the mean log-return $  \bar{\sigma}_k^2 T $ in both states $ k = 1,2 $. Equivalently, it means that the variance of the mean $  \bar{\sigma}_k^2 T $
is \emph{smaller} than the individual variance $ \sigma_k^2 T $  - which is as expected from the central limit theorem.   

\subsection{How far are we from the critical value $ h = h_c $ of the volatility parameter?}
\label{sect_far_to_critical_value}

If we use the values $ \bar{\rho}_M = 0.3-0.4 $ and $ N = 500 $, then Eqs.(\ref{A_and_g}) imply that $ A $ should approach one from below, and $ g \sim h^2/ (2 \bar{\sigma}_M^2 T) $.
On the other hand, using Eq.(\ref{A_i}), we can write the expression for $ A $ as follows:
\beq
\label{A_i_2}
A =  \frac{ h^2 (h^2 + h_c^2 - h^2) + 2 g h^2 \sigma^2 T + g^2 \sigma^4 T^2}{ \left(h^2 + g \sigma^2 T \right)^2} = 
1 - \frac{h^2 (h^2 - h_c^2)}{\left(h^2 + g \sigma^2 T \right)^2}  
\eeq
Given that $ A $ should be slightly below one, this implies that $ h^2 $ should be slightly above the critical 
value $ h_c^2 $. This means that the model is in a high-temperature phase, yet close to the bifurcation point $ h^2 = h_c^2 $.
In this regime, the non-vanishing mean field $ m $ is solely due to the explicit breaking of $ \mathbb{Z}_2 $ symmetry due to asymmetry of the potential,
which can be mimicked by adding the fictitious field $ B_0 $, see Eq.(\ref{B_0}). 

This behavior of the model appears to be quite reasonable. Indeed, a benign market regime is typically associated with a steadily growing market which, while exhibiting some volatility, has a positive trend. While it is not easy to estimate this trend accurately, a market with a positive market trend is certainly very different from a market with a negative trend. In a world with a symmetric potential and a second-order phase transition, a non-vanishing expected market return can only arise due to a spontaneous breaking of the $ \mathbb{Z}_2 $ symmetry, which implies that both choices are `physically' equivalent, i.e. equally preferable, in contradiction with the common sense. This suggests that the scenario in which the $ \mathbb{Z}_2 $ symmetry is broken \emph{explicitly} due to asymmetry of the potential, while the sign of the the mean field $ m $ is fixed by the potential, appears more plausible in the present context. The next section shows how the mean field $ m $ can be computed for this scenario using the MANES model calibrated to option prices.

\subsection{Expected market return from self-consistency equation}
\label{sect_market_return_from_self_consistency}

%In this section, I will assume that $ \bar{\mu}_1 = - \bar{\mu}_2 = \bar{\mu} $. 
%Perhaps the most practically important consequence of the theory developed in this paper is  a new relation for the equilibrium expected market log-return $ m $.
As was discussed above, the equilibrium expected market log-return is  
given by the mean field parameter $ m $ in the McKean-Vlasov equation (\ref{McKean_Vlasov}), and it 
should satisfy the self-consistency equation  (\ref{self_consist_NES}). 
Here I present an explicit solution of this equation, assuming that the model is calibrated to option prices, so that renormalized parameters
$   \bar{\mu}_1,  \bar{\mu}_2,  \bar{\sigma}_1, \bar{\sigma}_2, a , h $ are known.

%This still leave us with some freedom to impose additional constraints on the model parameters. 

Using Eq.(\ref{sigma_from_sigma_hat})
in Eq.(\ref{renormalized_params_NES_2}), the relations between the `bare' and `renormalized' parameters, resp. $ \mu_k $ and $ \bar{\mu}_k $ can be written as follows:
\beq
\label{mu_bar_from_mu_2}
\bar{\mu}_1 = \mu_1 \left( 1 - \frac{g}{h^2} \bar{\sigma}_1^2 T \right) + \frac{g}{h^2} \bar{\sigma}_1^2 m, \; \; \; 
\bar{\mu}_2 = \mu_2 \left( 1 - \frac{g}{h^2} \bar{\sigma}_2^2 T \right) + \frac{g}{h^2} \bar{\sigma}_2^2 m
\eeq 
According to Eq.(\ref{third_component}), the mean of the third Gaussian component in Eq.(\ref{three_component_Psi_2}) is as follows:
\beq
 \label{third_component_2}  
 \bar{\mu}_3 = 
% \frac{ \bar{\sigma}_2^2 \left( 1 - \frac{g}{h^2} \bar{\sigma}_1^2 T \right) \mu_1 + 
%\bar{\sigma}_1^2 \left( 1 - \frac{g}{h^2} \bar{\sigma}_2^2 T \right) \mu_2 +
%\frac{2g}{h^2} \bar{\sigma}_1^2 \bar{\sigma}_2^2 m}{
%\bar{\sigma}_1^2 + \bar{\sigma}_2^2} =  
\frac{ \bar{\mu}_1 \bar{\sigma}_2^2}{\bar{\sigma}_1^2 + \bar{\sigma}_2^2} + 
 \frac{ \bar{\mu}_2 \bar{\sigma}_1^2}{\bar{\sigma}_1^2 + \bar{\sigma}_2^2}
\eeq
% , \; \; \; 
%  \frac{\sigma_3^2}{2}  =   \frac{ \sigma_1^2 \sigma_2^2  }{\sigma_1^2 + \sigma_2^2}
% The normalization condition thus fixes the value of the constant $ C $ as follows:
% \beq
% \label{C_2}
% C^2 =  \frac{2 \sqrt{\pi T}}{ \Omega}, \; \; \; \text{where} \; \; \; \Omega =  \frac{(1-a)^2}{\sigma_1} + 
%  \frac{a^2}{\sigma_2}  + 
%  \frac{ 2 a(1-a)}{\sqrt{(\sigma_1^2 + \sigma_2^2)/2}} e^{ - \frac{ (\mu_1 - \mu_2)^2 T}{2(\sigma_1^2 + \sigma_2^2)} } 
%  \eeq 
 The self-consistency condition  (\ref{self_consist_NES}) is equivalent to the constraint on the expected value of $ y $ obtained in the model with the effective potential $ V_{eff} $ with the interaction-dressed parameters defined in Eqs.(\ref{mu_bar_from_mu_2}) and (\ref{third_component_2}).
 Therefore, the self-consistency condition  (\ref{self_consist_NES}) now takes the following simple form:
 \beq
 \label{self_consist_GM}
 m = \bar{\omega}_1 \bar{\mu}_1 T + \bar{\omega}_2 \bar{\mu}_2 T + \bar{\omega}_3 \bar{\mu}_3 T 
 \eeq
 where weights $ \bar{\omega}_k $ are defined as in Eq.(\ref{GM_weights}) using the `interaction-dressed' parameters $ \bar{\mu}_1, \bar{\mu}_2, \bar{\sigma}_1, \bar{\sigma}_2, h, \bar{a} $. Using here Eq.(\ref{third_component_2})
 and re-grouping terms, we obtain
% \bea
% \label{self_consist_GM_2}
% && \left( 1 - \frac{g}{h^2} \sigma_1^2 T \right)  \left( \omega_1  + \frac{\omega_3 \bar{\sigma}_2^2}{ \bar{\sigma}_1^2 + \bar{\sigma}_2^2}  \right) \mu_1 T +
%  \left( 1 - \frac{g}{h^2} \sigma_2^2 T \right)  \left( \omega_2  + \frac{\omega_3 \bar{\sigma}_1^2}{ \bar{\sigma}_1^2 + \bar{\sigma}_2^2}  \right) \mu_2 T 
% \nonumber \\ 
% && + 
%\left( -1 + \frac{g}{h^2} \left( \omega_1 \bar{\sigma}_1^2 + \omega_2 \bar{\sigma}_2^2 + 2 \omega_3 \frac{ \bar{\sigma}_1^2 \bar{\sigma}_2^2}{
%\bar{\sigma}_1^2 + \bar{\sigma}_2^2} \right) \right) m = 0
% \eea
% 
% Plugging into this relations Eqs.(\ref{renormalized_params_NES_2}) for $ \bar{\mu}_1 $ and $ \bar{\mu}_2 $ and re-grouping terms, one obtains
 \beq
 \label{m_answer_GM}
% m =  \frac{  \left( \bar{\omega}_1  + \bar{\omega}_3 \frac{\bar{\sigma}_2^2}{\bar{\sigma}_1^2 + \bar{\sigma}_2^2} \right)   \bar{\mu}_1 T
% + \left( \bar{\omega}_2  + \bar{\omega}_3 \frac{\bar{\sigma}_1^2}{\bar{\sigma}_1^2 + \bar{\sigma}_2^2} \right)   \bar{\mu}_2 T
%}{
%1 - \frac{g}{h^2} \left( \bar{\omega}_1 \bar{\sigma}_1^2 T + \bar{\omega}_2 \bar{\sigma}_2^2 T \right)}
m = \left( \bar{\omega}_1  +  \frac{\bar{\omega}_3 \bar{\sigma}_2^2}{\bar{\sigma}_1^2 + \bar{\sigma}_2^2} \right)   \bar{\mu}_1 T
 + \left( \bar{\omega}_2  +  \frac{\bar{\omega}_3 \bar{\sigma}_1^2}{\bar{\sigma}_1^2 + \bar{\sigma}_2^2} \right)   \bar{\mu}_2 T
\eeq      
This relation provides a closed-form solution of the self-consistency condition (\ref{self_consist_NES}) 
in terms of renormalized parameters that are directly calibrated to 
option prices. 
Recall that Eqs.(\ref{renormalized_params_NES}) imply that renormalized weights $ \bar{\omega}_k $ depend on `bare' weights $ \omega_k $ and the mean field $ m $.  Therefore, have we used the bare weights $ \omega_k $ instead of renormalized weights $ \bar{\omega}_k $ and expressed $ \bar{\mu}_1, \, \bar{\mu}_2 $
in terms of $ \mu_1, \, \mu_2 $ and $ m $ according to Eq.(\ref{mu_bar_from_mu_2}), 
Eq.(\ref{m_answer_GM}) would amount to a non-linear equation similar to the self-consistency equation (\ref{self_consist_NES_4}). 
Instead, by working with the MANES effective potential (\ref{pot_V_eff_3}) and renormalized parameters (\ref{mu_bar_from_mu_2}), the original self-consistency condition (\ref{self_consist}) of the McKean-Vlasov equation (\ref{McKean_Vlasov}) is resolved here analytically rather than numerically. 
Interestingly, when expressed in terms of renormalized parameters in Eq.(\ref{m_answer_GM}), the expected market log-return $ m $ does not explicitly depend
on the coupling constant $ g $, and is found in terms of parameters directly calibrated to option prices. 

On the other hand, the `bare' model parameters $ \mu_1, \mu_2 $ etc. \emph{do} depend on the value of $ g $. In particular, 
after the mean field $ m $ is computed from Eq.(\ref{m_answer_GM}), the `bare' model parameters $ \mu_1, \mu_2 $ can be obtained from Eq.(\ref{mu_bar_from_mu_2}), volatilities $ \sigma_1, \sigma_2 $ are obtained using Eq.(\ref{sigma_from_sigma_hat}), and the bare mixing coefficient $ a $ can be computed by inverting the formula for $ \bar{a} $ in Eq.(\ref{renormalized_params_NES}):
 \beq
 \label{bar_a}
 a = \left[ 1 + \frac{1 - \bar{a}}{ \bar{a}} \sqrt{ \frac{ h^2 + g \sigma_1^2T}{ h^2 + g \sigma_2^2 T}} e^{ 
 \frac{g (m-\mu_1 T)^2}{2( h^2 + g \sigma_1^2 T)} -  
 \frac{ g (m-\mu_2 T)^2}{2(h^2 + g \sigma_2^2 T)} 
 } \right]^{-1}  
\eeq 
 
%\section{Practical implementation and applications to option pricing}
%\label{sect_Applications}
%
\subsection{Examples of calibration to SPX options}
\label{sect_numerical_examples}

%\subsection{Examples of calibration to SPX options}
%\label{sect_calib_to_SPX_options}

In this section, I present three sets of examples of calibration to market quotes on European options on SPX (the S\&P 500 index). 
I use the same set of option quotes that were used in \cite{NES} to illustrate the working of the NES model. 
In all three sets of experiments, I calibrate to market quotes on 10 call options and 10 put options.
The strikes are chosen among available market quotes to cover the the range of option deltas between 0.02 and 0.5, in absolute terms, so that the model is 
calibrated 
to both ATM strikes and deep OTM strikes. Details of the loss function are described in \cite{NES}.
Optimization of the loss function is done using the \verb|shgo| algorithm available in the Python scientific computing package scipy.

In all experiments presented below, parameters $ \bar{\mu}_1, \bar{\mu}_2, \bar{\sigma}_1, \bar{\sigma}_2, \bar{a}, h $ are found by calibration
to SPX options. Inferred parameters are different from those reported in \cite{NES}, because here I do \emph{not} enforce the  
constraint $ \bar{\mu}_2 = - \bar{\mu}_1 $
that was used in \cite{NES}.
% which results in inferred model parameters $ \bar{\mu}_1, \bar{\mu}_2, \bar{\sigma}_1, \bar{\sigma}_2, \bar{a}, h $
%that are different from values found in \cite{NES}.
In addition, I display estimated parameters $ g $ and $ m $.
The coupling constant $ g $ is roughly estimated according to the following formula that is obtained by combining Eqs.(\ref{A_and_g}) and (\ref{constr_sigma_M}):
\beq
\label{g_estimated}
g \simeq \frac{h^2}{2 \bar{\sigma}_M^2 T}, \; \; \; \bar{\sigma}_M^2 = \max_{k} \left(  \frac{\bar{\sigma}_k^2}{2} \right) + \Delta \sigma^2
\eeq    
where $ \Delta \sigma^2 $ is a margin that controls the strength of the inequality (\ref{constr_sigma_M}). For numerical examples, the value $ \Delta \sigma^2 = 0.05 $ will be used in the examples below. Furthermore, the equilibrium log-return (mean field) $ m $ is computed according to Eq.(\ref{m_answer_GM}).
% where
%the coupling constant $ g $ is estimated using Eq.(\ref{g_estimated}). As was explained above, the chosen values of $ g $ and $ m $ do not impact parameters
%$ \bar{\mu}_1, \bar{\mu}_2, \bar{\sigma}_1, \bar{\sigma}_2, \bar{a}, h $ which are obtained by calibration to option prices. This implies that varying the values 
%of $ g $ and $ m $ would not change the quality of calibration to options.  

\begin{figure}[]
\begin{center}
\includegraphics[
width=130mm,
height=55mm]{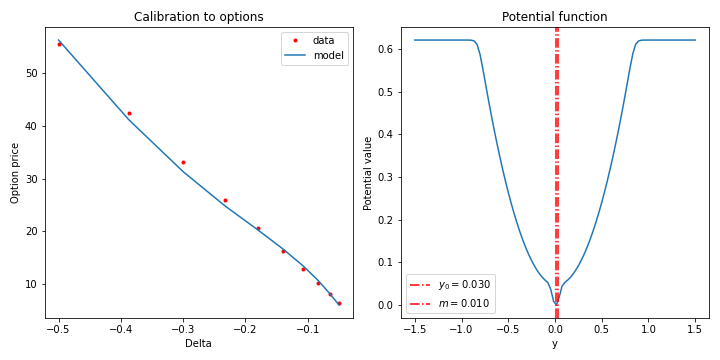}
\caption{Calibration to 1M SPX put options with expiry 08/09/2021 on 07/12/2021and the corresponding effective potentials (\ref{pot_V_eff_3}).  The vertical red lines correspond to the current value of log-return $ y_0 $ and the equilibrium log-return $ m $.
} 
\label{calibration_SPX_1M_puts}
\end{center}
\end{figure}

\begin{figure}[]
\begin{center}
\includegraphics[
width=130mm,
height=55mm]{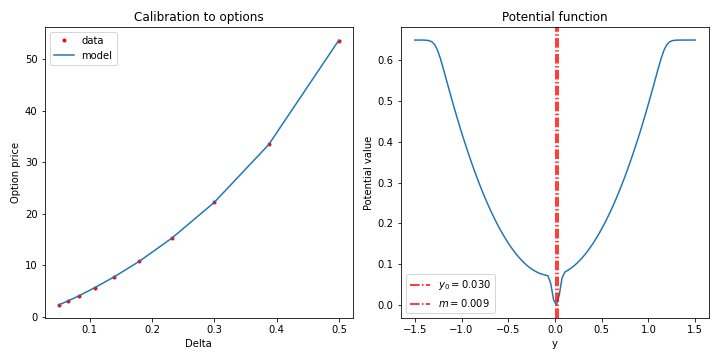}
\caption{Calibration to 1M SPX call options with expiry 08/09/2021 on 07/12/2021 and the corresponding effective potentials (\ref{pot_V_eff_3}).  The vertical red lines correspond to the current value of log-return $ y_0 $ and the equilibrium log-return $ m $.
} 
\label{calibration_SPX_1M_calls}
\end{center}
\end{figure}

In the first example, I consider SPX 1M options on 07/12/2021 with maturity on 08/09/2021. The results of calibration are presented in Table \ref{tab_NES_params_1M_SPX} and Figs.~\ref{calibration_SPX_1M_puts} and \ref{calibration_SPX_1M_calls}. Potentials shown in these figures and 
in the examples to follow are effective potentials (\ref{pot_V_eff_3}), and are computed as a single-stock NES potential with parameters 
$ \bar{\mu}_1, \bar{\mu}_2, \bar{\sigma}_1, \bar{\sigma}_2, \bar{a}, h $ inferred from calibration to index options.
Note the difference in implied potentials for puts and calls, as well as different values of inferred model parameters. For both potentials, the current value of the log-return $ y_0 $, as shown by the vertical red lines, 
is located near the bottom of the potential well. Furthermore, the equilibrium log-return $ m $ is also located near the bottom of the potential for for potentials.
This suggests that the price dynamics on this date correspond to an equilibrium regime of small fluctuations around a stable minimum.  

\begin{table}
\begin{center}
  \begin{tabular}{lSSSSSSSSS}
    \toprule
%    \multirow{2}{*}{Parameter} &
%      \multicolumn{5}{c}{Calibration to call options} &
%      \multicolumn{5}{c}{Calibration to put options}  \\
%      & {$ \mu $} & {$ \sigma_1 $} & {$ \sigma_2 $} &  {$ a $} & {$ h $}   & { \text{MAPE} } \\
%      \midrule
%    Puts &0.092 & 0.09 &0.461 &0.505 & 0.159   & 0.035\\
%    Calls &0.191 & 0.07 & 0.263 &0.566 & 0.162   & 0.002 \\
    & {$ \bar{\mu}_1 $} & {$ \bar{\mu}_2$ } & {$ \bar{\sigma}_1 $} & {$ \bar{\sigma}_2 $} &  {$ \bar{a} $} & {$ h $}  & {$g $} & {$ m $}  & { \text{MAPE} } \\
      \midrule
    Puts &0.190 & 0.118  & 0.449  & 0.093  & 0.525 & 0.159 & 1.23  & 0.01 & 3.45\%\\
    Calls &0.136 & -0.592 & 0.089 & 0.669 & 0.342 & 0.163 &  0.63 & 0.01 & 0.293\% \\

    \bottomrule
  \end{tabular}
 \caption{NES parameters obtained by calibration to 10 put and 10 call options on 1M SPX options with expiry 08/09/2021 on 07/12/2021.
 The last column shows the mean absolute pricing errors (MAPE).} 
\label{tab_NES_params_1M_SPX} 
\end{center}  
\end{table}


\begin{figure}[]
\begin{center}
\includegraphics[
width=130mm,
height=55mm]{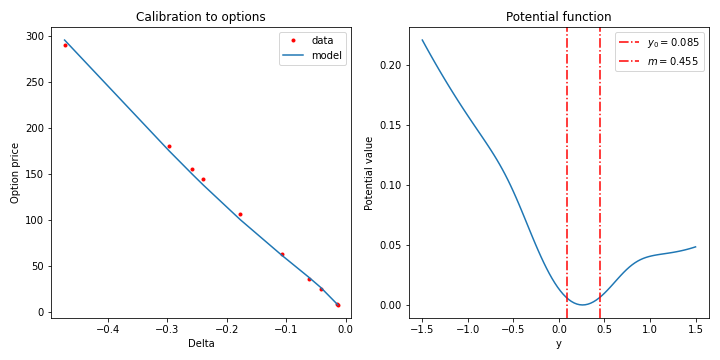}
\caption{Calibration to 1Y SPX put options with expiry 09/21/2021 on 11/06/2020 and the corresponding effective potentials (\ref{pot_V_eff_3}). The vertical red lines correspond to the current value of log-return $ y_0 $ and the equilibrium log-return $ m $.
} 
\label{calibration_SPX_1Y_puts}
\end{center}
\end{figure}

\begin{figure}[]
\begin{center}
\includegraphics[
width=130mm,
height=55mm]{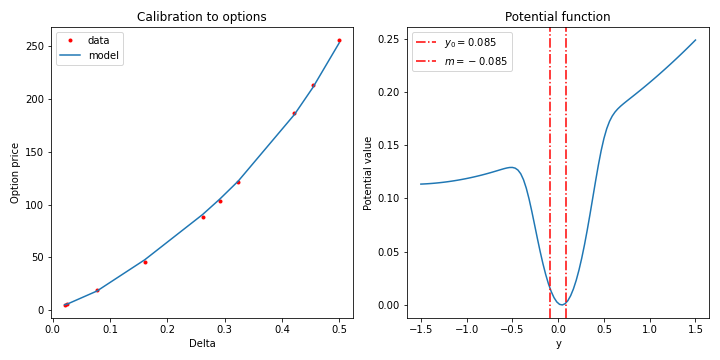}
\caption{Calibration to 1Y SPX call options with expiry  09/21/2021 on 11/06/2020  and the corresponding effective potentials (\ref{pot_V_eff_3}).  The vertical red lines correspond to the current value of log-return $ y_0 $ and the equilibrium log-return $ m $.
} 
\label{calibration_SPX_1Y_calls}
\end{center}
\end{figure}

In the second example, I look at longer option tenors, and consider SPX 1Y options on 11/06/2020 with maturity on 09/21/2021. The results of calibration are presented in Table \ref{tab_NES_params_1Y_SPX} and Figs.~\ref{calibration_SPX_1Y_puts} and \ref{calibration_SPX_1Y_calls}. Again, we can note the difference in implied effective potentials for puts and calls, as well as different values of inferred model parameters. Also as in the previous example, for both potentials, the current value of the log-return $ y_0 $ and equilibrium expected log-return $ m $ are located near the bottom of the potential wells,  suggesting an equilibrium regime of small price fluctuations on this date.  

\begin{table}
\begin{center}
  \begin{tabular}{lSSSSSSSSS}
    \toprule
%    \multirow{2}{*}{Parameter} &
%      \multicolumn{5}{c}{Calibration to call options} &
%      \multicolumn{5}{c}{Calibration to put options}  \\
%      & {$ \mu $} & {$ \sigma_1 $} & {$ \sigma_2 $} &  {$ a $} & {$ h $}   & { \text{MAPE} } \\
%      \midrule
%    Puts &0.092 & 0.251 &0.813 &0.405 & 0.165   & 0.055\\
%    Calls &0.106 & 0.123 & 0.505 &0.565 & 0.217   & 0.019 \\
   & {$ \bar{\mu}_1 $} & {$ \bar{\mu}_2$ } & {$ \bar{\sigma}_1 $} & {$ \bar{\sigma}_2 $} &  {$ \bar{a} $} & {$ h $}  & {$g $} & {$ m $}  & { \text{MAPE} } \\
      \midrule
    Puts &1.200 & 0.269  & 0.906 &0.303  & 0.438 & 0.198 & 0.05 & 0.455 & 3.49\%\\
    Calls &0.047 & -1.866 & 0.168 & 1.423 & 0.467 & 0.221 &  0.03 & - 0.085 & 2.80\% \\
     
    \bottomrule
  \end{tabular}
 \caption{NES parameters obtained by calibration to 10 put and 10 call options on 1Y SPX options with expiry 09/21/2021 on 11/06/2020. The last column shows the mean absolute pricing errors (MAPE).} 
\label{tab_NES_params_1Y_SPX} 
\end{center}  
\end{table}


\begin{figure}[]
\begin{center}
\includegraphics[
width=130mm,
height=55mm]{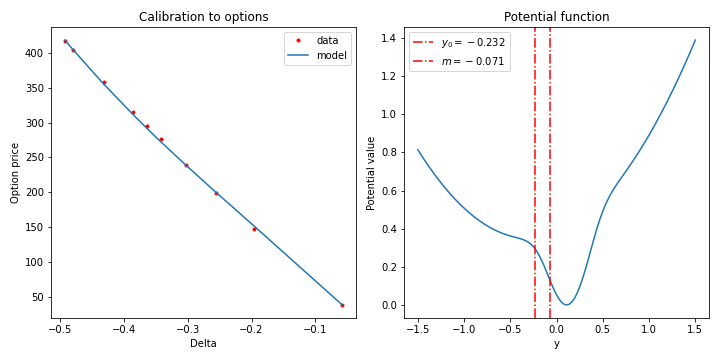}
\caption{Calibration to 6M SPX put options with expiry  09/18/2020 on 03/16/2020 and the corresponding effective potentials (\ref{pot_V_eff_3}). The vertical red lines correspond to the current value of log-return $ y_0 $ and the equilibrium log-return $ m $.
} 
\label{calibration_SPX_6M_puts}
\end{center}
\end{figure}

\begin{figure}[]
\begin{center}
\includegraphics[
width=130mm,
height=55mm]{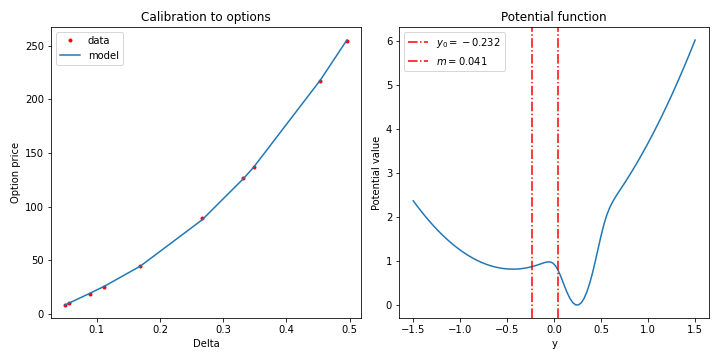}
\caption{Calibration to 6M SPX call options with expiry  09/18/2020 on 03/16/2020 and the corresponding effective potentials (\ref{pot_V_eff_3}).  The vertical red lines correspond to the current value of log-return $ y_0 $ and the equilibrium log-return $ m $.
} 
\label{calibration_SPX_6M_calls}
\end{center}
\end{figure}

Finally, the last example considers 6M options with the expiry on 09/18/2020 on 03/16/2020, at the peak of Covid-19 crisis where the SPX index had the largest drop. This example thus illustrates the model behavior for a severely distressed market. 
The results of calibration are presented in Table \ref{tab_NES_params_6M_SPX} and Figs.~\ref{calibration_SPX_6M_puts} and \ref{calibration_SPX_6M_calls}. As in the previous examples, again note the difference in implied effective potentials for puts and calls, as well as different values of inferred model parameters. 

\begin{table}
\begin{center}
  \begin{tabular}{lSSSSSSSSS}
    \toprule
%    \multirow{2}{*}{Parameter} &
%      \multicolumn{5}{c}{Calibration to call options} &
%      \multicolumn{5}{c}{Calibration to put options}  \\
   & {$ \bar{\mu}_1 $} & {$ \bar{\mu}_2$ } & {$ \bar{\sigma}_1 $} & {$ \bar{\sigma}_2 $} &  {$ \bar{a} $} & {$ h $}  & {$g $} & {$ m $}  & { \text{MAPE} } \\
      \midrule
%    Puts &0.579 & 0.200   & 0.386 &0.959  & 0.438 & 0.556 & 0.57 & 0.316  & 0.592\%\\    
%    Calls &0.527 & -1.430 & 0.165 & 0.260 &0.083 & 0.827 & 12.27 & 0.352 & 1.99\% \\   
       Puts &0.234 & -0.585   & 0.229 &1.10  & 0.757 & 0.630 & 0.57 & -0.07  & 0.956\%\\
%      Calls & 0.526 & -1.965 & 0.165  & 1.386 & 0.006 & 0.827 & 0.63 & 0.27 & 1.995\% \\  
      Calls & 0.487 & -0.864 & 0.146  & 0.700 & 0.624 & 0.831 & 2.28 & 0.04 & 1.664\% \\      		
    \bottomrule
  \end{tabular}
 \caption{NES parameters obtained by calibration to 10 put and 10 call options on 6M  SPX options with expiry  09/18/2020 on 03/16/2020. The last column shows the mean absolute pricing errors (MAPE).} 
\label{tab_NES_params_6M_SPX} 
\end{center}  
\end{table}

\emph{Unlike} the previous examples, for the present case of a distressed market, 
the initial log-return $ y_0 $ on 03/16/2020 is located far away from the global minima for both effective potentials, 
indicating that this initial state is strongly non-equilibrium, consistently with a prevailing market sentiment on that date.  
%strongly non-equilibrium dynamics on that date.  Both sets of put and call market quotes indicate that the value of log-return observed on  03/16/2020 is a strongly 
%non-equilibrium initial state, .

Interestingly, while both implied effective potentials suggest that the current value $ y_0 $ is far from the global minimum of the potential and 
hence describes a non-equilibrium scenario, they differ in the character of a subsequent relaxation mechanism for this initial state. The implied potential for puts in 
 Fig.~\ref{calibration_SPX_6M_puts} 
is a single-well potential, therefore the initial state $ y_0 $ is unstable. If the potential itself remains constant through time, this initial state would eventually relax into its global minimum. This would happen even in the limit of zero volatility (zero noise). When the noise is present with $ h > 0 $, it produces  
an uncertainty for the time needed to reach the minimum of the potential, as well as small fluctuations around this minimum.

Differently from the puts, the call options seem to suggest a different type of relaxation dynamics, as the implied potential for the calls in   Fig.~\ref{calibration_SPX_6M_calls} is a double well potential quite similar to the one shown in the left column in Fig.~\ref{fig_IQED_WF_and_potentials}.
The initial location is in the vicinity of the local minimum corresponding to the left well, while the right well corresponds to the global minimum. As they are separated by the barrier, the potential in Fig.~\ref{calibration_SPX_6M_calls} describes a scenario of metastability. The relaxation to the global minimum (the true ground state) proceeds via instanton transitions as discussed in \cite{HD_QED, NES}. Instanton transitions are only possibly
when the volatility parameter is non-zero, $ h > 0 $, however small it can be in practice. 
This suggests that double well potentials implying metastable dynamics may occur when markets are in distress or  
during periods of a high market uncertainty, e.g. during general crises such as the 2008 crisis or the Covid-19 crisis of 2020, or during general elections.\footnote{ 
In particular, \cite{Gemmill} found a bimodal implied distribution during British elections of 1987 (though not through other British elections in 1992), and suggested that option prices can be used to monitor the market sentiment during elections.}

\section{Summary and outlook}
\label{sect_Summary}

%McCauley in his insightful and provocative book \cite{McCauley}  pointed out various problems with traditional equilibrium approaches of classical financial models such as e.g. the CAPM model \cite{CAPM} or the Black-Scholes model \cite{BS}. In particular, he argued that fat tails in financial data, which is a matter of everyday concerns of practitioners and academics alike, may be simply the result of applying an unjustified binning process to non-stationary data.
%McCauley argued against a neoclassical economic doctrine based on the concept of a market-clearing equilibrium, and showed that among about five different definitions of a market equilibrium commonly used in the mathematical financial modeling literature, none makes sense from the physics' perspective (see also 
%\cite{Inverted_World} for related arguments). He also issued a challenge to econophysics as an emerging rival of the classical finance, where research often explicitly or implicitly assumes an equilibrium, potentially suffering from the same problem of a potential data distortion leading to the above-mentioned problem with problematically measured fat tails. 

This paper proposed a tractable non-linear model of interacting and non-equilibrium market, formulated as statistical mechanics of 
interacting non-linear oscillators where individual stocks' log-returns $ y_i $ are viewed as coordinates of `particles' describing these stocks.  
Both self-interactions of oscillators 
$ y_i $ and their pairwise interactions are explained in terms of money flows into the market from external investors. In particular, correlations between log-returns of individual stocks originate from the dependence of money flow $ a_i $ into stock $ i $ on the average previous performance of all other stocks. 
  
The main idea of this paper was to approximate the return of a market index in a heterogeneous market by
a mean field of a \emph{homogeneous} market made of $ N $ replicas of the same `representative' stock with a self-interaction potential $ V(y) $. This enables 
modeling the dynamics of the market log-return using the mean field approximation that produces the McKean-Vlasov equation as the equation governing the 
dynamics of the system. As the McKean-Vlasov equation is a non-linear equation corresponding to the thermodynamic limit $ N \rightarrow \infty $, it 
produces far richer dynamics than the original multi-particle Fokker-Planck equation (\ref{FPE}), giving rise to ergodicity breaking and first- and second-order phase transitions in different parameter regimes. The resulting dynamics of the mean field resembles the Desai-Zwanzig model of interacting non-linear oscillators
\cite{Desai_Zwanzig_1978, Dawson_1983} and its generalized version in \cite{Gomes_2019}.  
 
Furthermore, while the exploration of the phase structure in the generalized Desai-Zwanzig model requires dedicated numerical methods to 
identify stable and unstable solutions \cite{Gomes_2019}, the analysis in the present paper is considerably simplified due to its reliance on the  
 Non-Equilibrium Skew (NES) potential (\ref{V_Psi_0}). The NES model with the NES potential (\ref{V_Psi_0}) was introduced in \cite{NES}
 as a flexible and highly tractable non-linear model of single-stock dynamics that is capable of describing either a benign or stressed market regime,
 and tracks the pre-asymptotic dynamic behavior of transition probabilities. In particular, non-equilibrium, pre-asymptotic corrections to the moments of a stationary (asymptotic) distribution of log-returns are explicitly controlled in the NES model in terms of the model parameters entering the NES potential (\ref{V_Psi_0}). 
 
 In the present paper, the NES potential from \cite{NES} is used to model a multi-asset market.
 In Sect.~\ref{sect_MANES_potential}, the NES potential is introduced as a tractable 
 approximation to a non-linear self-interaction potential $ V(y) $  that originates in money flows and their impact effect on market returns.
 In addition, the quadratic Curie-Weiss interaction potential (\ref{pair_interaction_pot}) is obtained as an approximate interaction potential following the same lines of analysis. This results in a multi-asset extension of the previous single-stock NES model, referred to as the Multi-Asset NES (MANES) model in this paper.
 
 As was shown in this paper, the new multi-asset MANES model is as tractable as the previous single-stock NES model. This is made possible due to the fact 
 that with the NES self-interaction potential $ V(y) $, the new effective potential $ V_{eff}(y) $ that incorporates interactions in the system can be expressed in terms of a single-stock NES potential $ V(y) $ with renormalized parameters, see  Eqs.(\ref{pot_V_eff_3}) and (\ref{renormalized_params_NES}). This suggests that for the purpose of calibration to market prices of index options, the multi-asset MANES model is computationally equivalent to a single-stock NES model applied to the `representative' stock. Also due to this property of the model, the self-consistency equation
 of the mean field approximation that is normally expressed as a non-linear equation (\ref{self_consist}) is resolved here analytically in terms of renormalized 
 parameters that are directly calibrated to market prices of index options. Furthermore, estimates made in Sect.~\ref{sect_far_to_critical_value} suggest that 
 the model operates in a `high-temperature' phase close to the criticality point, where the volatility parameter $ h $ is slightly higher than the critical value 
 $ h = h_c $ that corresponds to the bifurcation point of the order parameter $ m $.
  
Just as the single-stock NES model, the MANES model 
demonstrates that a \emph{single volatility parameter} is sufficient to accurately match available market prices of index options. 
 These results stand in stark contrast to the most of other option pricing models such as local, stochastic, or rough volatility models that need more complex specifications of noise to fit the market data. Alongside a single volatility parameter $ h $, 
 the effective 
 single-stock potential $ V_{eff}(y) $ (\ref{pot_V_eff_3}) has parameters $ \bar{\mu}_1, \bar{\mu}_2, \bar{\sigma}_1, \bar{\sigma}_2, \bar{a} $ that can be 
 calibrated to market prices of index options. 
 By fitting these parameters to the market data, we produce implied potentials which replace 
 implied volatility smiles as a way to fit market data. If desired, the calibrated model parameters $ \bar{\mu}_1, \bar{\mu}_2, \bar{\sigma}_1, \bar{\sigma}_2, \bar{a} $
 can be approximately mapped onto the parameters that enter Eq.(\ref{U_self_potential}), and thus could be interpreted as as implied money flow and market impact parameters. Clearly, the success of this approach for matching market prices of vanilla index options does not preclude one from using more sophisticated models of noise such as stochastic or rough volatility models - but only if needed beyond the need to explain market prices of vanilla options.  
 
%The approach of this paper offers a viable alternative mechanism based on the analysis of market flows and their impact. As shown in \cite{HD_QED} and discussed in this paper, the interplay of these effects creates a non-linear drift potential in the stock price dynamics. The NES model presented in this paper provides a parameterized model of these dynamics, with drift parameters $  \mu, \sigma_1, \sigma_2, a $ that 
% define the drift potential, and a single volatility parameter $ h $.  
          
The MANES model can be used for several applications or practical interest. 
In particular, option-implied moments of future returns can be used as predictors for actual future returns, volatilities and skewness, and employed for portfolio trading, see e.g. \cite{Gemmill, Stilger}. While such analyses typically use
risk-neutral moments implied by option prices, the MANES model enables extracting both risk-neutral and real-measure moments, and thus enriches the set of predictors for such tasks. The model can also use other market data for model calibration. In particular, in addition to using market prices of options, we 
could incorporate open interest data in the model calibration. The MANES model could also be 
jointly calibrated to the equity and credit markets data, by adding a fit to credit indices such as CDX as 
proxies to probabilities of large market drops. 
Such applications and extensions are left here for a future research.     
%
%For calibration of the NES model to single-stock options, one could consider a similar approach that would use single-name credit default swaps (CDS) instead of credit indices. Another interesting direction of research is to develop a portfolio optimization framework with single stock dynamics as specified in this paper. 

%The model developed in this paper can be used and extended in a number of interesting ways. In particular, initial numerical experiments suggest that risk-neutral measures separately implied by calls or put options are typically different, contrary to the traditional lore of the mathematical finance literature. The latter typically takes the existence of a unique risk-neutral measure for granted, and then estimates it in empirical research by a joint calibration to both calls and puts, making its approach somewhat circular. Exploring optimal combinations of information inferred separately from call and put options, and possibly incorporating new signals, e.g. credit indexes such as CDX could be an interesting future direction.  Another potential direction would be to analyze 
%applications of this framework to individual stocks, rather than to a market index. While formulas derived in this paper could also be formally applied to individual stocks, their individual dynamics should be consistent with the dynamics of the market as a whole. Such extensions are left here for a future research.      

%\clearpage
\appendix

\def\thesection{A}	
\setcounter{equation}{0}
\def\theequation{\thesection.\arabic{equation}}

%\section*{Appendix A: Instantons in the Langevin dynamics}
%\label{sect:Appendix_A}
%
%\def\thesection{B}
%\setcounter{equation}{0}
%\def\theequation{\thesection\arabic{equation}}
%
%\section*{Appendix B: SUSY}
%\label{sect_Appendix_SUSY}

%\clearpage
 
\end{document}